\documentclass[a4paper,11pt]{article}
\usepackage{jheppub} 
\usepackage[T1]{fontenc}
\usepackage{booktabs}
\usepackage[dvips,table]{xcolor} 
\usepackage{xspace}
\usepackage{dcolumn}
\usepackage[subrefformat=parens,position=top,skip=-10pt,margin=0pt,justification=justified,singlelinecheck=false]{subcaption}

\newcommand{\Pl}{\ell}
\newcommand{\fb}{{\ensuremath\unskip\,\text{fb}}\xspace}
\def\reffi#1{\mbox{Figure~\ref{#1}}}
\def\reffis#1{\mbox{Figures~\ref{#1}}}
\def\refta#1{\mbox{Table~\ref{#1}}}

\def\refse#1{\mbox{Section~\ref{#1}}}
\def\citere#1{\mbox{Ref.~\cite{#1}}}
\def\citeres#1{\mbox{Refs.~\cite{#1}}}

\newcommand{\ri}{\mathrm i}

\def\be{\begin{equation}}
\def\ee{\end{equation}}

\newcommand{\qqb}{\ensuremath{q\bar{q}}\xspace}
\newcommand{\PH}{\ensuremath{\text{H}}\xspace}
\newcommand{\Pj}{\ensuremath{\text{j}}\xspace}
\newcommand{\Pp}{\ensuremath{\text{p}}\xspace}

\newcommand{\Pb}{\ensuremath{\text{b}}\xspace}
\newcommand{\Pq}{\ensuremath{\text{q}}\xspace}
\newcommand{\Pt}{\ensuremath{\text{t}}\xspace}
\newcommand{\Pu}{\ensuremath{\text{u}}\xspace}
\newcommand{\Pd}{\ensuremath{\text{d}}\xspace}
\newcommand{\Ps}{\ensuremath{\text{s}}\xspace}
\newcommand{\Pc}{\ensuremath{\text{c}}\xspace}
\newcommand{\Pg}{\ensuremath{\text{g}}\xspace}

\newcommand{\PW}{\ensuremath{\text{W}}\xspace}
\newcommand{\PZ}{\ensuremath{\text{Z}}\xspace}

\newcommand{\Mt}{\ensuremath{m_\Pt}\xspace}
\newcommand{\MH}{\ensuremath{M_\PH}\xspace}
\newcommand{\MWOS}{\ensuremath{M_\PW^\text{OS}}\xspace}
\newcommand{\MW}{\ensuremath{M_\PW}\xspace}
\newcommand{\MZOS}{\ensuremath{M_\PZ^\text{OS}}\xspace}
\newcommand{\MZ}{\ensuremath{M_\PZ}\xspace}
\newcommand{\Mb}{\ensuremath{m_\Pb}\xspace}
\newcommand{\Gt}{\ensuremath{\Gamma_\Pt}\xspace}
\newcommand{\GH}{\ensuremath{\Gamma_\PH}\xspace}

\newcommand{\GZOS}{\ensuremath{\Gamma_\PZ^\text{OS}}\xspace}
\newcommand{\GW}{\ensuremath{\Gamma_\PW}\xspace}
\newcommand{\GWOS}{\ensuremath{\Gamma_\PW^\text{OS}}\xspace}

\newcommand{\GeV}{\ensuremath{\,\text{GeV}}\xspace}
\newcommand{\TeV}{\ensuremath{\,\text{TeV}}\xspace}

\newcommand{\alphas}{\ensuremath{\alpha_\text{s}}\xspace}
\newcommand{\order}[1]{\ensuremath{\mathcal{O}{\left(#1\right)}}\xspace}

\newcommand{\deltar}{\ensuremath{\Delta R}\xspace}

\newcommand{\GF}{\ensuremath{G_\mu}}

\newcommand{\ph}{p_\text{H}}
\newcommand{\phath}{\hat{p}_\text{H}}
\newcommand{\pt}{p_\text{t}}
\newcommand{\phatt}{\hat{p}_\text{t}}
\newcommand{\ptx}{p_{\bar{\text{t}}}}
\newcommand{\phattx}{\hat{p}_{\bar{\text{t}}}}
\newcommand{\phatwplus}{\hat{p}_{\text{W}^+}}
\newcommand{\phatwminus}{\hat{p}_{\text{W}^-}}

\newcommand{\pbtprime}{p_{\text{b}_{\text{t}}}^\prime}
\newcommand{\phatbt}{\hat{p}_{\text{b}_{\text{t}}}}

\newcommand{\pbxtxprime}{p_{\bar{\text{b}}_{\bar{\text{t}}}}^\prime}
\newcommand{\pwplus}{p_{\text{W}^+}}
\newcommand{\pwminus}{p_{\text{W}^-}}

\newcommand{\ttbarbbbar}{\ensuremath{\Pt\bar{\Pt}\Pb\bar{\Pb}}\xspace}
\newcommand{\ttbarh}{\ensuremath{\Pt\bar{\Pt}\PH}\xspace}
\newcommand{\ttbarhProcess}{\ensuremath{\Pp\Pp\to\Pt\bar{\Pt}\PH\to\Pl^+\nu_\Pl \Pj\Pj\Pb\bar{\Pb}\Pb\bar{\Pb}}\xspace}
\newcommand{\ttbarProcess }{\ensuremath{\Pp\Pp\to\Pt\bar{\Pt}\Pb\bar{\Pb}\to\Pl^+\nu_\Pl \Pj\Pj\Pb\bar{\Pb}\Pb\bar{\Pb}}\xspace}
\newcommand{\fullProcess  }{\ensuremath{\Pp\Pp\to\Pl^+\nu_\Pl \Pj\Pj\Pb\bar{\Pb}\Pb\bar{\Pb}}\xspace}
\newcommand{\ggProcess    }{\ensuremath{\Pg\Pg\to\Pl^+\nu_\Pl q'\bar{q}''\Pb\bar{\Pb}\Pb\bar{\Pb}}\xspace}
\newcommand{\qqbProcess    }{\ensuremath{q\bar{q}\to\Pl^+\nu_\Pl q'\bar{q}''\Pb\bar{\Pb}\Pb\bar{\Pb}}\xspace}
\newcommand{\ttbbb}{\ttbarbbbar}

\def\M12{\ensuremath M_{\mathrm{b_1 b_2}}}
\def\MdR{\ensuremath M_{\mathrm{bb},\Delta R_\mathrm{min}}}
\def\MHP{\ensuremath M_{{\mathrm {bb, Higgs}}}}
\def\Mttb{\ensuremath M_{\mathrm{bb, non\textnormal{-}top}}}

\newcolumntype{.}{D{.}{.}{-1}}
\newcolumntype{d}[1]{D{.}{.}{#1}}

\colorlet{tableoverheadcolor}{gray!37.5}
\colorlet{tableheadcolor}{gray!25}
\colorlet{tablerowcolor}{gray!12.5}

\newcommand{\gsim}
{\;\raisebox{-.3em}{$\stackrel{\displaystyle >}{\sim}$}\;}

\title{Irreducible background and interference effects for 
Higgs-boson production in association with a top-quark pair}
\subheader{\today}

\author{A. Denner,}
\author{R. Feger,}
\author{A. Scharf}

\affiliation{%
        Universit\"at W\"urzburg, %
        Institut f\"ur Theoretische Physik und Astrophysik, %
        Emil-Hilb-Weg 22, \linebreak %
        97074 W\"urzburg, %
        Germany%
}
\emailAdd{ansgar.denner@physik.uni-wuerzburg.de}
\emailAdd{robert.feger@physik.uni-wuerzburg.de}
\emailAdd{andreas.scharf@physik.uni-wuerzburg.de}

\abstract{We present an analysis of Higgs-boson production in
  association with a top-quark pair at the LHC investigating in
  particular the final state consisting of four b jets, two jets, one
  identified charged lepton and missing energy. We consider the
  Standard Model prediction in three scenarios, the resonant
  Higgs-boson plus top-quark-pair production, the resonant production
  of a top-quark pair in association with a b-jet pair and the full
  process including all non-resonant and interference contributions.
  By comparing these scenarios we examine the irreducible background
  for the production rate and several kinematical distributions. With
  standard selection criteria the irreducible background turns out to
  be three times as large as the signal. For most observables we find
  a uniform deviation of eight percent between the scenario requiring
  two resonant top quarks and the full process. In particular
  phase-space regions the non-resonant contributions cause larger
  effects, and we observe shape changes for some distributions.
  Furthermore we investigate interference effects 
  and find that neglecting interference contributions results in an
  over-estimate of the total cross-section of five percent.}
\toccontinuoustrue \notoc

\begin{document} 

\maketitle
\flushbottom

\section{Introduction}
\label{sec:introduction}

The discovery of the long-sought Higgs boson at the LHC in July 2012 
\cite{Aad:2012tfa,Chatrchyan:2012ufa} ushered in a new era of probing
the mechanism of spontaneous symmetry breaking and thus mass
generation in nature.  The determination of the properties of the
discovered Higgs boson is a major goal of the LHC. Especially the
couplings of the Higgs boson to matter particles are important for the
understanding of the origin of mass. The production of a Higgs boson
in association with a top-quark pair is of particular interest, since
it allows to directly access the top-quark Yukawa coupling. Although
the production rate is small and the measurement is experimentally
challenging, ATLAS \cite{ATLAS:2012cpa,ATLAS:2014,Aad:2014lma} 
and CMS \cite{CMS:2012qaa,CMS:2013sea,CMS-PAS-HIG-13-015,Chatrchyan:2013yea,Khachatryan:2014qaa}
have already performed searches using the data of the LHC runs at 7
and $8\TeV$. With the upcoming run 2 of the LHC at $13\TeV$ the
determination of the $\ttbarh$ signal and the potential measurement of
the top-quark Yukawa coupling will be pursued.

The production of a Higgs boson in association with a top--antitop
pair has been studied theoretically by many authors.  Leading-order
(LO) predictions for the production of $\Pt\bar\Pt\PH$ for stable
Higgs boson and top quarks have been presented in
\citeres{Raitio:1978pt,Ng:1983jm,Kunszt:1984ri,Gunion:1991kg,Marciano:1991qq}
while the corresponding next-to-leading-order (NLO) corrections have
been calculated in
\citeres{Beenakker:2001rj,Beenakker:2002nc,Reina:2001sf,Dawson:2002tg,
  Dawson:2003zu}. More recently the matching of the NLO corrections to
parton showers has been performed
\cite{Frederix:2011zi,Garzelli:2011vp}. Very recently also electroweak
corrections to $\Pt\bar\Pt\PH$ have been elaborated
\cite{Frixione:2014qaa,Yu:2014cka}. NLO QCD corrections for the most
important irreducible background process, $\Pt\bar\Pt\Pb\bar\Pb$
production in LO QCD, have been worked out
\cite{Bredenstein:2008zb,Bredenstein:2009aj,Bevilacqua:2009zn,Bredenstein:2010rs}
and matched to parton showers
\cite{Kardos:2013vxa,Cascioli:2013era,TROCSANYI:2014lha}. A combined
analysis of $\Pt\bar\Pt\PH$ and $\Pt\bar\Pt\Pb\bar\Pb$ production was
carried out in \citere{Binoth:2010ra}. Further analyses and results
can be found in the yellow reports of the LHC Higgs Cross Section
Working Group
\cite{Dittmaier:2011ti,Dittmaier:2012vm,Heinemeyer:2013tqa}.

In this article we study Higgs-boson production in association with a
top-quark pair ($\ttbarh$) including the subsequent semileptonic decay
of the top-quark pair and the decay of the Higgs boson into a
bottom--antibottom-quark pair,
\begin{equation}
\ttbarhProcess.
\end{equation}
The final state under consideration consists of six jets, four of
which are b jets, one identified charged lepton (electron or muon) and
missing energy, and our primary goal is the study of the irreducible
background. We consider this process in three different scenarios. In
the first scenario we include the complete Standard Model (SM)
contributions, comprising all resonant, non-resonant and interference
contributions to the 8-particle final state. For the second scenario
we require the intermediate resonant production of a top-quark pair in
association with a bottom-quark pair. We employ the pole approximation
for the top quarks and include the leptonic/hadronic decay of the
top/antitop quark. Similar approaches have been used for $\Pt\bar{\Pt}$
production in association with massless particles at NLO QCD
\cite{Melnikov:2011ta,Melnikov:2011qx}. In the third scenario,
corresponding to the signal, we require in addition to the resonant
top-quark pair an intermediate resonant Higgs boson decaying into a
bottom-quark pair and employ the pole approximation for the Higgs
boson as well.  We use the matrix-element generator
RECOLA~\cite{Actis:2012qn} to compute all matrix elements at leading
order in perturbation theory.

Comparing the predictions in the three scenarios allows us to
examine the size of the irreducible background for Higgs production in
association with a top--antitop-quark pair.  Further we determine the
quality of the approximations compared to the calculation of the full
process and quantify deviations for the total cross section and
differential distributions.  We investigate the total cross section
and differential distributions for the LHC operating at $13\TeV$ with
particular emphasis on distributions that allow to enhance the signal
over the irreducible background.  We study, in particular, different
methods of assigning a b-jet pair to the Higgs boson and compare their
performance in reconstructing the Higgs signal. Furthermore, we
investigate the size of interference effects between contributions to
the matrix elements of different order in the strong and electroweak
coupling constants.

The paper is organised as follows. In Section~\ref{sec:calculation} we
specify the setup of our calculation and identify the various partonic
contributions to the signal process. In
Section~\ref{sec:results} we present numerical results for the total
cross section and kinematical distributions and quantify
the size of the irreducible background.  Our conclusions are
presented in Section~\ref{sec:conclusion}. In
Appendix~\ref{sec:OnShellProjection} we explain in detail the on-shell
projections needed for the pole approximations we apply and in
Appendix~\ref{sec:FurtherObs} we present additional results.

\section{Notation and setup}
\label{sec:calculation}

This section provides some technical details about our computation. We
consider gluons and light (anti-)quarks ($\Pu, \Pd, \Pc, \Ps$) to be
the only constituents of the proton and disregard contributions from
bottom quarks and photons in the parton distribution functions.  We
neglect flavour mixing as well as finite-mass effects for the light
quarks and leptons. The considered matrix elements involve
(multiple) resonances of electroweak gauge bosons, top quarks and the
Higgs boson. For the consistent description of these resonances we use
the complex-mass scheme
\cite{Denner:1999gp,Denner:2005fg,Denner:2006ic} where all masses of
unstable particles are defined by the poles of the propagators in the
complex plane, $\mu^2=m^2-\ri m\Gamma$. In addition, all couplings,
and, in particular the weak mixing angle, are consistently derived from
the complex masses and thus are complex, too. 

We consider three scenarios to calculate the process $\fullProcess$:
\begin{itemize}
\item In the first scenario, the \textit{full process}, we include all SM
  contributions to the process $\fullProcess$. Counting $\Pu$, $\Pd$,
  $\Pc$, and $\Ps$ quarks separately, we distinguish 48~partonic
  channels. Eight channels can be constructed from
  $\ggProcess$ and 40 from $\qqbProcess$ using crossing symmetries and
  substituting different quark flavours.  Matrix elements involving
  external gluons receive contributions of $\order{\alphas\alpha^3}$,
  $\order{\alphas^2\alpha^2}$ and $\order{\alphas^3\alpha}$, whereas
  amplitudes without external gluons receive an additional
  $\order{\alpha^4}$ term of pure electroweak origin.  Some sample
  diagrams are shown in
  \reffis{fig:full_process_feynman_diagrams_1}--\ref{fig:full_process_feynman_diagrams_3}.
\begin{figure}
        \setlength{\parskip}{-10pt}
        \begin{subfigure}{0.32\linewidth}
                \subcaption{}
                \includegraphics[width=\linewidth]{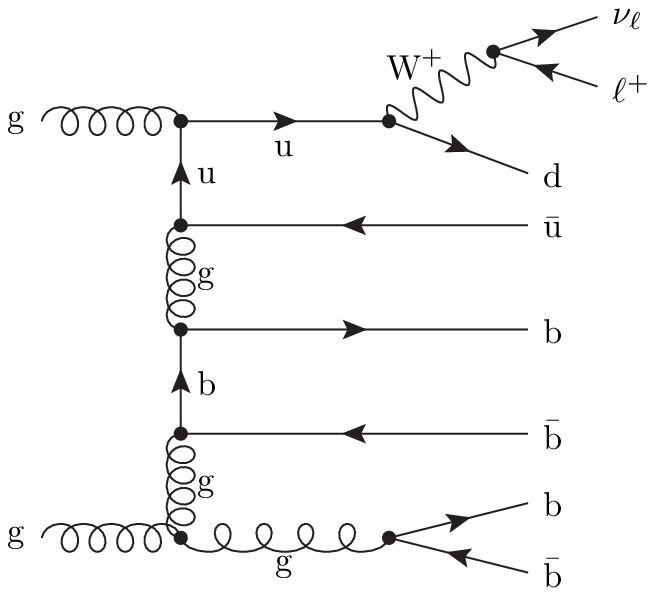}
                \label{fig:full_process_feynman_diagrams_1} 
        \end{subfigure}
        \begin{subfigure}{0.32\linewidth}
                \subcaption{}
                \includegraphics[width=\linewidth]{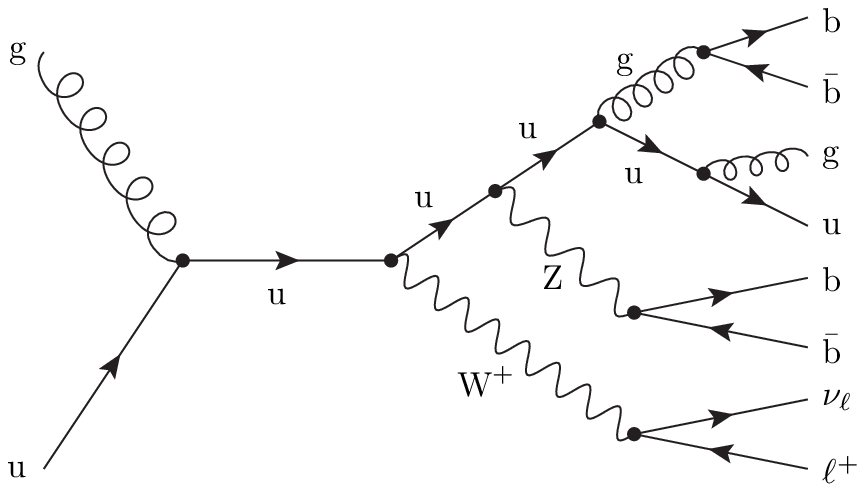}
                \label{fig:full_process_feynman_diagrams_2} 
        \end{subfigure}
        \hfill
        \begin{subfigure}{0.32\linewidth}
                \subcaption{}
                \includegraphics[width=\linewidth]{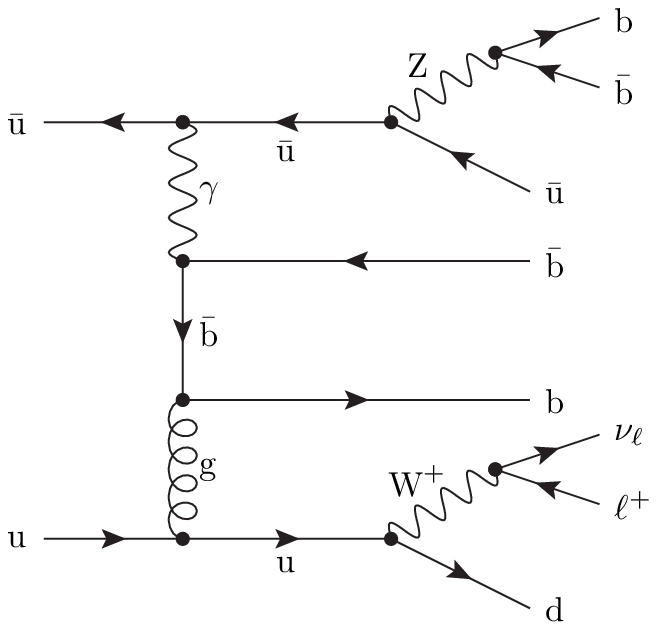}
                \label{fig:full_process_feynman_diagrams_3}
        \end{subfigure}
        
        \begin{subfigure}{0.32\linewidth}
                \subcaption{}
                \includegraphics[width=\linewidth]{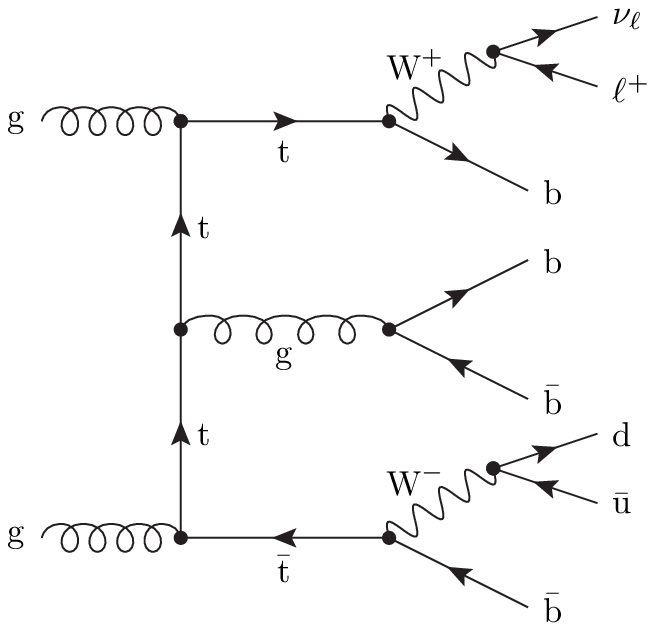}
                \label{fig:ttbarbbbar_feynman_diagrams_1} 
        \end{subfigure}
        \begin{subfigure}{0.32\linewidth}
                \subcaption{}
                \includegraphics[width=\linewidth]{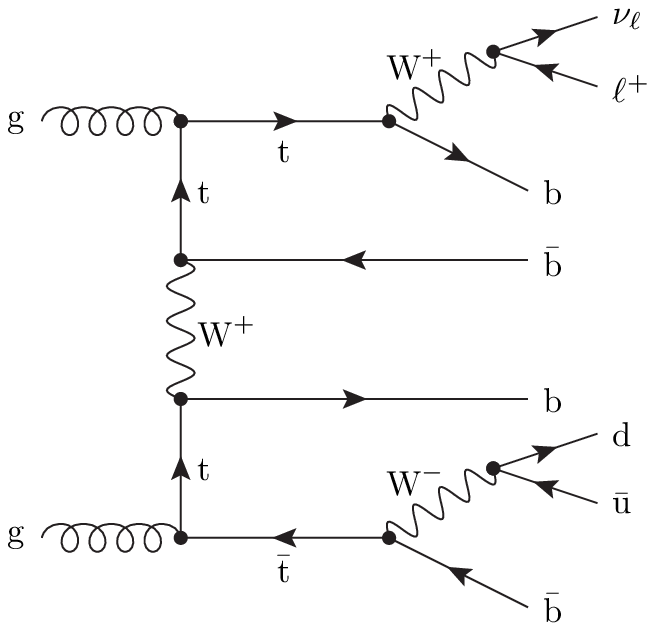}
                \label{fig:ttbarbbbar_feynman_diagrams_2} 
        \end{subfigure}
        \hfill
        \begin{subfigure}{0.32\linewidth}
                \subcaption{}
                \includegraphics[width=\linewidth]{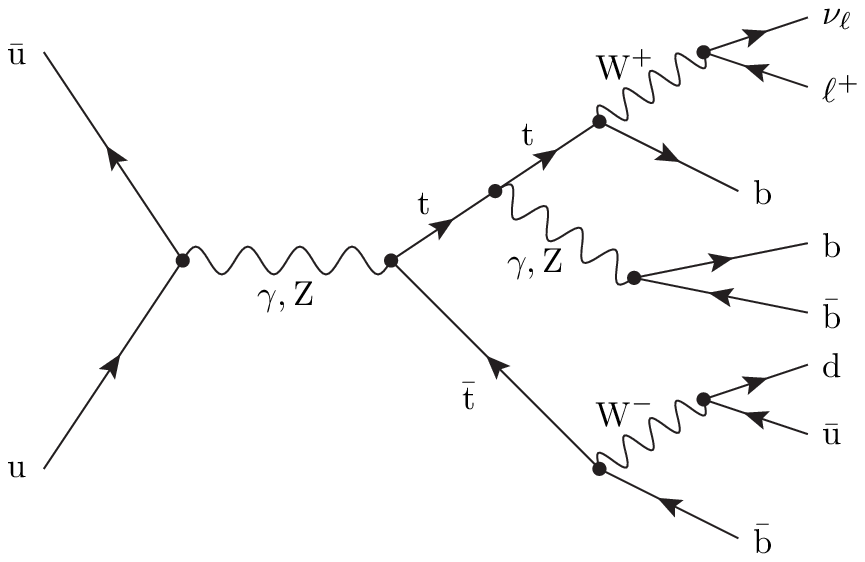}
                \label{fig:ttbarbbbar_feynman_diagrams_3} 
        \end{subfigure}
        
        \begin{subfigure}{0.32\linewidth}
                \subcaption{}
                \includegraphics[width=\linewidth]{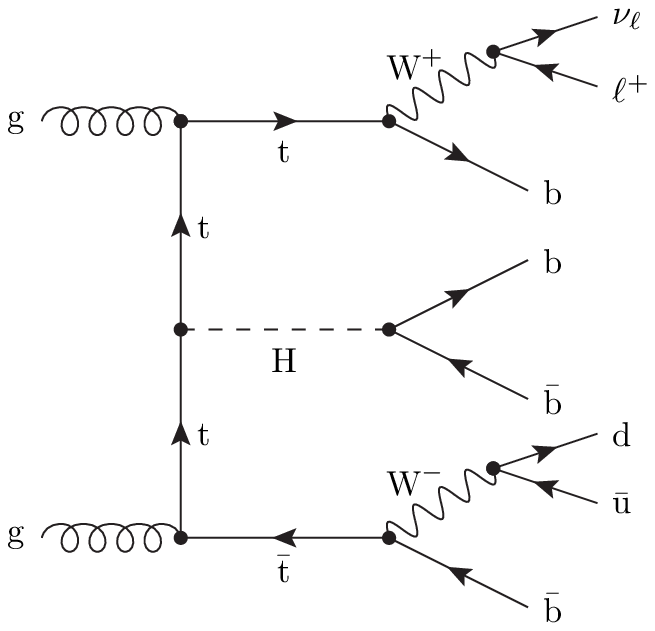}
                \label{fig:ttbarh_feynman_diagrams_1} 
        \end{subfigure}
        \begin{subfigure}{0.32\linewidth}
                \subcaption{}
                \includegraphics[width=\linewidth]{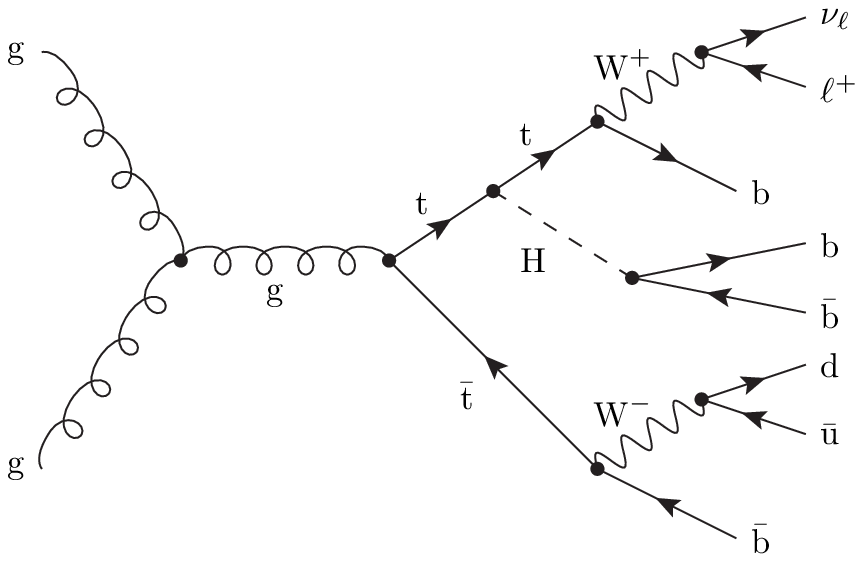}
                \label{fig:ttbarh_feynman_diagrams_2} 
        \end{subfigure}
        \hfill
        \begin{subfigure}{0.32\linewidth}
                \subcaption{}
                \includegraphics[width=\linewidth]{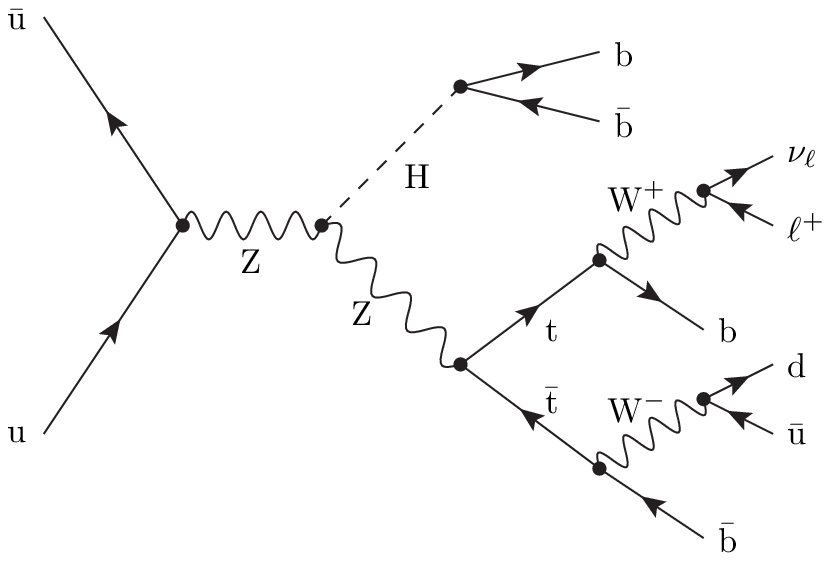}
        \label{fig:ttbarh_feynman_diagrams_3} 
        \end{subfigure}%

        \caption{\label{fig:feynman_diagrams}%
                Representative Feynman diagrams for %
                \subref{fig:full_process_feynman_diagrams_1}--\subref{fig:full_process_feynman_diagrams_3} the full process (top), %
                \subref{fig:ttbarbbbar_feynman_diagrams_1}--\subref{fig:ttbarbbbar_feynman_diagrams_3} $\ttbarbbbar$ production (middle) and %
                \subref{fig:ttbarh_feynman_diagrams_1}--\subref{fig:ttbarh_feynman_diagrams_3} $\ttbarh$ production (bottom).%
        }
\end{figure}
\item In the second scenario we only take those diagrams into account
  that contain an intermediate top--antitop-quark pair. The resulting
  amplitude, labelled \textit{$\ttbarbbbar$ production} in the
  following, corresponds to the production of a bottom--antibottom
  pair and an intermediate top--antitop pair followed by its
  semileptonic decay, i.e. $\ttbarProcess$. Some sample diagrams are
  shown in
  \reffis{fig:ttbarbbbar_feynman_diagrams_1}--\ref{fig:ttbarbbbar_feynman_diagrams_3}.
  Note that we use the pole approximation for the top quarks only,
  hence we take into account all off-shell effects of the remaining
  unstable particles. The details of our implementation of the pole
  approximation are described in Appendix~\ref{sec:OnShellProjection}.
  This scenario involves 10 partonic channels, comprising 2 gluon-fusion and
  8 quark--antiquark-annihilation channels.  As a consequence of the
  required top--antitop-quark pair the amplitudes receive no
  contribution of $\order{\alphas^3\alpha}$.
\item Finally, we consider the signal process $\ttbarhProcess$ and
  label it \textit{$\ttbarh$ production}. In addition to the
  intermediate top--antitop-quark pair we require an intermediate
  Higgs boson decaying into a bottom--antibottom-quark pair and use
  the pole approximation for the top-quark pair and the Higgs boson.
  Here the same 10 partonic channels as in the previous case
  contribute.  The requirement of the Higgs boson eliminates
  contributions of the $\order{\alphas^2\alpha^2}$ from the amplitude.
  The implementation of the pole approximation applied to the Higgs
  boson is explained in Appendix~\ref{sec:OnShellProjection}, and some
  sample diagrams are shown in
  \reffis{fig:ttbarh_feynman_diagrams_1}--\ref{fig:ttbarh_feynman_diagrams_3}.
\end{itemize}

The full process involves partonic channels with up to 78,000
diagrams. All matrix elements are calculated with
RECOLA~\cite{Actis:2012qn} which provides a fast and numerically
stable computation. RECOLA computes the matrix elements using
recursive methods, i.e.\ the complexity does not scale with the number
of Feynman diagrams, and allows to specify intermediate particles for
a given process. The phase-space integration is performed with an
in-house multi-channel Monte-Carlo program. Here the number of
diagrams matters for the construction of the integration channels.

\section{Results}
\label{sec:results}

We present results for the LHC operating at $13\TeV$. For the
calculation of the hadronic cross section we employ LHAPDF 6.0.5 with
the CT10 LO parton distributions (LHAPDF ID 10800). Following
\citere{Beenakker:2002nc} we use a fixed renormalisation and
factorisation scale
\begin{equation}\label{eqn:FixedScale}
        \mu = \mu_\text{R} = \mu_\text{F} = \frac{1}{2}\left(2m_t + m_H\right) = 236\GeV.
\end{equation}
The corresponding value of the strong 
coupling constant $\alphas$ provided by LHAPDF reads
\begin{equation}
        \alphas(\mu) = 0.103219\ldots,
\end{equation}
based on a running at one-loop accuracy and $N_\text{F}=5$ active
flavours. We neglect any contribution from the suppressed bottom-quark
parton density.

For the masses and widths of unstable particles we use
\begin{equation}\label{eqn:ParticleProperties}
        \begin{aligned}
                \Mt   &= 173\GeV,           & \Gt   &= 1.47\GeV,\\
                \MH   &= 126\GeV,           & \GH   &= 4.21\times 10^{-3}\GeV,\\
                \MZOS &=  91.1876\GeV,\quad & \GZOS &= 2.4952\GeV,\\
                \MWOS &=  80.385\GeV,       & \GWOS &= 2.0850\GeV,\\
                \Mb   &=   4.8\GeV,
        \end{aligned}
\end{equation}
\newcommand{\MVOS}{\ensuremath{M_V^\text{OS}}\xspace}%
\newcommand{\GVOS}{\ensuremath{\Gamma_V^\text{OS}}\xspace}%
where the masses \MVOS and widths \GVOS of the \PZ and \PW bosons
($V=\PW,\PZ$) are the measured on-shell values (OS), which we convert
into pole values according to \citere{Bardin:1988xt}
\begin{equation}
        M_V = \MVOS/\sqrt{1+(\GVOS/\MVOS)^2}\,,\qquad  \Gamma_V = \GVOS/\sqrt{1+(\GVOS/\MVOS)^2}.
\end{equation}
All other masses and widths in \eqref{eqn:ParticleProperties} are
considered as pole values.

The width of the top quark \Gt in \eqref{eqn:ParticleProperties} for
unstable W bosons is computed at leading order according to
\citere{Jezabek:1988iv}, i.e.\ 
\newcommand{\gammaw}{\ensuremath{\gamma_\text{W}}}
\begin{equation}
        \Gamma_t^\text{LO} = \frac{\GF\Mt^5}{16\sqrt{2}\pi^2\MW^2}\int_0^{(1-\epsilon)^2}\!\!
                \frac{\text{d}y\,\gammaw}{\left(1-y/\bar{y}\right)^2+\gammaw^2 } F_0(\epsilon,y)
\end{equation}
with $\epsilon=\Mb/\Mt$, $\gammaw = \GW/\MW$, $\bar{y} = (\MW/\Mt)^2$ and
\begin{align}
        F_0(\epsilon, y)     &=  \frac{1}{2}\sqrt{\lambda(\epsilon, y)} f_0(\epsilon, y),         \nonumber\\
        f_0(\epsilon, y)     &=  4 \left[(1 - \epsilon^2)^2 + y (1 + \epsilon^2) - 2 y^2\right], \nonumber\\
        \lambda(\epsilon, y) &=  1 + y^2 + \epsilon^4 - 2 (y + \epsilon^2 + y \epsilon^2).
\end{align}
The masses and widths of all other quarks and leptons are neglected.

We derive the electromagnetic coupling $\alpha$ from the Fermi constant in the 
$G_\mu$ scheme \cite{Denner:2000bj}, 
\begin{equation}
  \alpha = \frac{\sqrt{2}}{\pi} G_\mu \MW^2 \left( 1 - \frac{\MW^2}{\MZ^2} \right),
  \qquad     \GF    = 1.16637\times 10^{-5}\GeV.            
\end{equation}

We impose cuts on the transverse momenta and rapidities of leptons,
jets, b~jets and missing transverse momentum as well as distance cuts
between all jets in the rapidity--azimuthal plane, with the distance
between two jets $i$ and $j$ defined as
\begin{equation}\label{eqn:DeltaR}
        \Delta R_{ij} = \sqrt{(\phi_i-\phi_j)^2+(y_i-y_j)^2},
\end{equation}
where $\phi_i$ and $y_i$ denote the azimuthal angle and rapidity of
jet $i$, respectively.

Our selection of cuts represents standard acceptance cuts and is neither 
deliberately chosen to enhance the contribution of \ttbarh production nor to 
suppress any background process:
\begin{equation}\label{eqn:cuts}
        \begin{aligned}
                \text{non-b jets:}                        && p_{\text{T},\Pj}         &>  25\GeV,  & |y_\Pj| &< 2.5, \hspace{15ex}\\
                \text{b jets:}                      && p_{\text{T},\Pb}         &>  25\GeV,  & |y_\Pb| &< 2.5,              \\
                \text{charged lepton:}              && p_{\text{T},\Pl^+  }         &>  20\GeV,  & |y_{\Pl^+}|   &< 2.5,              \\
                \text{missing transverse momentum:} && p_{\text{T},\text{miss}} &>  20\GeV,                                        \\
                \text{jet--jet distance:}           && \Delta R_{\Pj\Pj}        &> 0.4,                                            \\
                \text{b-jet--b-jet distance:}       && \Delta R_{\Pb\Pb}        &> 0.4,                                            \\
                \text{jet--b-jet distance:}         && \Delta R_{\Pj\Pb}        &> 0.4.                               
        \end{aligned}
\end{equation}

\subsection{Total cross section}

In this section we analyse the total hadronic cross section for the
three scenarios and cuts defined above. In
\refta{table:results_ttxh_onshell_projected} we show the total cross
section for \ttbarh production and the corresponding contributions
resulting from quark--antiquark annihilation and gluon fusion.
\begin{table}
        \centering
        \renewcommand\arraystretch{1.2}
        \rowcolors{1}{}{tablerowcolor}
        \begin{tabular}{ld{1.9}d{2.7}d{1.7}}
                \toprule\rowcolor{tableheadcolor}       
                \textbf{pp}            
                & \multicolumn{3}{>{\columncolor{tableheadcolor}}l}{\textbf{Cross section (fb)}}\\\rowcolor{tableheadcolor}
                & \multicolumn{1}{>{\columncolor{tableheadcolor}}c}{\boldmath$\order{(\alpha^4)^2}$}%
                & \multicolumn{1}{>{\columncolor{tableheadcolor}}c}{\boldmath$\order{(\alphas\alpha^3)^2}$}%
                & \multicolumn{1}{>{\columncolor{tableheadcolor}}c}{\bf Total}\\
                \midrule
                $q \bar{q}$ & 0.014887(2)                                            & 2.1467(2)    &  2.1621(2) \\
                $\Pg \Pg$   & \multicolumn{1}{>{\columncolor{tablerowcolor}}l}{--} & 5.230(1)     &  5.2298(9) \\\midrule\rowcolor{tableheadcolor}  
                $\sum$      & 0.014887(2)                                            & 7.377(1)     &  7.3920(9) \\
                \bottomrule
        \end{tabular}
        \caption{\label{table:results_ttxh_onshell_projected} 
                Composition of the total cross section in fb for \ttbarh production at the 
                LHC at $13\TeV$.  In the first column the partonic initial states are listed. In 
                the second and third column we list the contributions resulting from the square 
                of matrix elements of specific orders in the strong and electroweak coupling.  
                The last column provides the total cross section. 
        }
\end{table}
In this scenario the total cross section is $\sigma^{\rm
  Total}_{\ttbarh} = 7.39\fb$, and about $70\,\%$ of the events originate
from the gluon-fusion process. While the bulk of the contributions
results from matrix elements of order $\order{\alphas\alpha^3}$,
quark--antiquark annihilation receives an additional small
contribution from pure electroweak interactions. Note that there are
no interferences between diagrams of $\order{\alpha^4}$ and
$\order{\alphas\alpha^3}$ in this scenario.

The cross section for $\ttbarbbbar$ production is presented in
\refta{table:results_ttxbbx_onshell_projected}.
\begin{table}
        \centering
        \renewcommand\arraystretch{1.2}
        \rowcolors{1}{}{tablerowcolor}
        \begin{tabular}{ld{1.10}d{2.7}d{2.7}d{1.7}d{1.7}}
                \toprule\rowcolor{tableheadcolor}       
                \textbf{pp}            
                & \multicolumn{5}{>{\columncolor{tableheadcolor}}l}{\textbf{Cross section (fb)}}\\\rowcolor{tableheadcolor}
                & \multicolumn{1}{>{\columncolor{tableheadcolor}}c}{\boldmath$\order{(\alpha^4)^2}$}%
                & \multicolumn{1}{>{\columncolor{tableheadcolor}}c}{\boldmath$\order{(\alphas\alpha^3)^2}$}%
                & \multicolumn{1}{>{\columncolor{tableheadcolor}}c}{\boldmath$\order{(\alphas^2\alpha^2)^2}$}%
                & \multicolumn{1}{>{\columncolor{tableheadcolor}}c}{\bf Sum}%
                & \multicolumn{1}{>{\columncolor{tableheadcolor}}c}{\bf Total}\\
                \midrule
                $q \bar{q}$ &  0.018134(6)                                          &  2.4932(9) & 0.9199(2) &   3.4312(9) &   3.4366(6) \\
                $\Pg \Pg$   &  \multicolumn{1}{>{\columncolor{tablerowcolor}}l}{--} &  7.818(4)  & 16.650(9)  &  24.47(1)   &  23.010(7)  \\\midrule\rowcolor{tableheadcolor} 
                $\sum$      &  0.018134(6)                                          & 10.311(4)  & 17.570(9)  &  27.90(1)   &  26.446(7)  \\
                \bottomrule
        \end{tabular}
        \caption{\label{table:results_ttxbbx_onshell_projected}  
                Composition of the total cross section in fb for
                $\ttbarbbbar$ production at the LHC at $13\TeV$. In
                the first column the partonic initial states are
                listed. In the second, third and fourth 
                column we list the contributions resulting from the square
                of matrix elements of specific orders in the strong
                and electroweak coupling.  
                The fifth column shows the sum of the columns two to
                four, while the last column provides the total cross 
                section incorporating all interference effects. 
        }
\end{table}
In column two to four we show the contributions to the cross section
resulting from matrix elements of specific orders in $\alphas$.  The
corresponding matrix elements include only Feynman diagrams that
originate exclusively from \textit{one specific order} in the strong
and electroweak coupling constant, e.g.\ results in column three are
exclusively build upon Feynman diagrams of $\order{\alphas\alpha^3}$
and do not include interferences between diagrams of
$\order{\alpha^4}$ and $\order{\alphas^2\alpha^2}$. 
The fifth column, labelled "Sum" represents the sum of the
contributions in columns  two to four and thus contains no
interferences between matrix elements of different orders. The total
cross section in the last column is calculated from the complete
matrix element for \ttbarProcess including all interferences.
For the total
cross section we find a significant enhancement of the production rate
compared to \ttbarh production, $\sigma^{\rm Total}_{\ttbarbbbar} =
26.45\fb$, and thus  the irreducible background
\begin{equation*}
        \sigma^{\rm Irred.}_{\ttbarbbbar} = \sigma^{\rm Total}_{\ttbarbbbar} - \sigma^{\rm Total}_{\ttbarh}  = 19.06\fb
\end{equation*}
which exceeds the \ttbarh signal by a factor of 2.6. Note that in this
definition of the irreducible background interference effects between
signal and background amplitudes are included. The major contribution
to the irreducible background ($87\,\%$) arises from gluon fusion at
$\order{(\alphas^2\alpha^2)^2}$ while for quark--antiquark
annihilation we find a relative contribution of about $5\,\%$ at this
order. A comparison of the results at $\order{(\alphas\alpha^3)^2}$
between Table~\ref{table:results_ttxh_onshell_projected} and
Table~\ref{table:results_ttxbbx_onshell_projected} shows a rise of the
cross section of $49\,\%$ ($16\,\%$) for the gluon-fusion
(quark--antiquark-annihilation) process. These enhancements of the
$\order{(\alphas\alpha^3)^2}$ contribution in the
$\ttbarbbbar$~scenario result from Feynman diagrams involving
electroweak interactions with $\PZ$~bosons, $\PW$~bosons and photons
(like in \reffi{fig:ttbarbbbar_feynman_diagrams_2}).

A comparison between the fifth (Sum) and sixth (Total) column in
\refta{table:results_ttxbbx_onshell_projected} allows a determination
of interference contributions between matrix elements of different
orders in the coupling constants. Neglecting those interference
effects results in an over-estimation of the cross section of about
$5\,\%$. The main interference contributions originate from the
gluon-fusion channel, where they reduce the cross section by $6\,\%$, while
the $\qqb$ channel is hardly affected.  We could trace the dominant
contribution to interferences of diagrams of $\order{\alphas\alpha^3}$
with W-boson exchange in the $t$-channel (like in
\reffi{fig:ttbarbbbar_feynman_diagrams_2}) with diagrams of
$\order{\alphas^2\alpha^2}$ that yield the dominant irreducible
background (like in \reffi{fig:ttbarbbbar_feynman_diagrams_1}).  We
confirmed the size and sign of these contributions by switching off
all other contributions in RECOLA. We note that these kinds of
interferences are absent in the $\qqb$ channel.  We also investigated
the interference of the signal process, i.e.\ all diagrams of order
$\order{\alphas\alpha^3}$ involving $s$-channel Higgs-exchange
diagrams (like in \reffis{fig:ttbarh_feynman_diagrams_1} and
\ref{fig:ttbarh_feynman_diagrams_2}) with the dominant irreducible
background of order $\order{\alphas^2\alpha^2}$ and found it to be
below one per cent.

\begin{table}
        \centering
        \renewcommand\arraystretch{1.2}
        \setlength{\tabcolsep}{5pt}
        \rowcolors{1}{}{tablerowcolor}
        \begin{tabular}{ld{1.9}d{2.7}d{1.7}d{1.8}d{1.7}d{2.6}d{2.6}}
                \toprule\rowcolor{tableheadcolor}       
                \textbf{pp}                 
                & \multicolumn{6}{>{\columncolor{tableheadcolor}}l}{\textbf{Cross section (fb)}}\\\rowcolor{tableheadcolor}
                & \multicolumn{1}{>{\columncolor{tableheadcolor}}c}{\boldmath$\order{(\alpha^4)^2}$}%
                & \multicolumn{1}{>{\columncolor{tableheadcolor}}c}{\boldmath$\order{(\alphas\alpha^3)^2}$}%
                & \multicolumn{1}{>{\columncolor{tableheadcolor}}c}{\boldmath$\order{(\alphas^2\alpha^2)^2}$}%
                & \multicolumn{1}{>{\columncolor{tableheadcolor}}c}{\boldmath$\order{(\alphas^3\alpha)^2}$}%
                & \multicolumn{1}{>{\columncolor{tableheadcolor}}c}{\bf Sum}%
                & \multicolumn{1}{>{\columncolor{tableheadcolor}}c}{\bf Total}\\ 
                \midrule
                $\Pg q$          & \multicolumn{1}{l}{--} &   0.231(4)  &   0.370(2)  & 0.365(1)   &   0.966 (4)  &   0.944 (9) \\
                $\Pg \bar{q}$    & \multicolumn{1}{>{\columncolor{tablerowcolor}}l}{--} &   0.0421(6) &   0.0679(3) & 0.0608(2)  &   0.1708(7)  &   0.167 (1) \\
                $q q^{(\prime)}$ &  0.001471(2)                                         &   0.0575(5) &   0.1106(2) & 0.07871(9) &   0.2483(6)  &   0.2478(8) \\
                $q \bar{q}$      &  0.01973(3)                                          &   2.531(6)  &   0.957(1)  & 0.00333(1) &   3.511 (6)  &   3.538 (4) \\
                $\Pg \Pg  $      & \multicolumn{1}{l}{--} &   8.01(2)   &  17.19(6)   & 0.00756(2) &  25.21  (6)   &  23.71 (6) \\\midrule\rowcolor{tableheadcolor}                                         
                $\sum$           &  0.02120(3)                                          &  10.87(2)   &  18.69(6)   & 0.516(2)   &  30.10  (6)   &  28.60 (6) \\
                \bottomrule
        \end{tabular}
        \caption{\label{table:results_full_matrixelement_summary} 
                Composition of the total cross section in fb for
                the full process at the LHC at $13\TeV$. In the first
                column the partonic initial states are listed, where
                $q q^{(\prime)}$ denotes pairs of quarks and/or
                antiquarks other than $q\bar{q}$. In the columns two to
                five we list the contributions resulting from the square
                of matrix elements of specific orders in the strong
                and electroweak coupling. The sixth column shows the
                sum of the columns two to five, while the last column
                provides the total cross section incorporating all
                interference effects. 
        }
\end{table}
The results for the full process are listed in
\refta{table:results_full_matrixelement_summary}. We show
contributions resulting from matrix elements of different orders
similar as in \refta{table:results_ttxbbx_onshell_projected}. In
addition we list the contributions of additional partonic channels
separately. Here $\Pg\Pq$ and $\Pg\bar\Pq$ denote channels with gluons
and quarks or antiquarks in the initial state and $qq^{(\prime)}$ all
channels involving two quarks and/or antiquarks in the initial state
other than $\qqb$ (including channels with gluons in the final state).
We compute the total cross section to $\sigma^{\rm Total}_{\rm full} =
28.60 \fb$ including all interference effects, with the major
contribution (about $83\,\%$) arising from the purely gluon-induced
process.  The consideration of contributions without an intermediate
top--antitop-quark pair results in an increase of about $3\,\%$ for
both the $\Pg\Pg$ and $\qqb$ processes compared to the $\ttbarbbbar$
scenario.  The additional partonic channels contribute about $5\,\%$.
For the total cross section we find a relative increase of $8\,\%$
while for the irreducible background
\begin{equation*}
        \sigma^{\rm Irred.}_{\rm full} = \sigma^{\rm Total}_{\rm full} - \sigma^{\rm Total}_{\ttbarh}  = 21.21 \fb
\end{equation*}
we note an enhancement of $11\,\%$ relative to $\ttbarbbbar$ production. 
The signal cross section $\sigma^{\rm Total}_{\ttbarh}$ amounts to
$26\,\%$ of the full cross section  $\sigma^{\rm Total}_{\rm full}$.

In the full process sizeable interference effects only appear in the
gluon-induced channel and are of the same size, roughly $-6\,\%$, as for
\ttbarbbbar production.  Since the total cross section of the full
process and \ttbarbbbar production differ by only about $8\,\%$ we
conclude that the major interference effects are those that we
identified in the underlying \ttbarbbbar production process.

\subsection{Differential distributions}

In this section we present differential distributions for all three
scenarios.  We compare results for the full process with $\ttbarbbbar$
production and $\ttbarh$ production to assess the irreducible
background to $\ttbarh$ production in distributions.  In the upper panels
in each plot we show the differential distribution of the full process
with a black solid line, \ttbarbbbar production with a dashed blue
line and \ttbarh production with a dotted red line. The lower panels
provide the ratio of \ttbarbbbar production to the full process with a
dashed blue line and the ratio of \ttbarh production to the full
process with a dotted red line.

We study specifically the distribution in the invariant mass of two
b jets selected according to different criteria in view of identifying
the b-jet pairs originating from the Higgs-boson decay. In particular, we
consider the invariant mass of
\begin{itemize}
        \item 
                the two b jets that form the invariant mass closest to the Higgs-boson mass, $\MHP$,
        \item 
                the two b jets that are most likely not originating from the (anti-)top-quark decay, $\Mttb$,
        \item 
                the two b jets that have the smallest $\Delta R$ distance as defined in \eqref{eqn:DeltaR}, $\MdR$,
        \item 
                the two b jets that have the highest transverse momenta, $\M12$.
\end{itemize}

Figure~\ref{fig:Mbb_one} displays the corresponding b-jet pair
invariant-mass distributions.  In all cases the shape of the full
process is well represented by the $\ttbbb$ approximation including
the Higgs-boson and the $\PZ$-boson resonance. In the lower panels we
observe that the difference between $\ttbbb$~production and the full
process is generally of the same order as for the total cross section
of about $8\,\%$. For \ttbarh production the shape and the relative size
compared to the full process depends on the observable. In the
following we analyse in detail the four invariant-mass distributions.

\subsubsection{Invariant mass of b-jet pair closest to the Higgs-boson mass}
\label{subsubsec:Best_Higgs_Mass_Method}
\begin{figure}
        \setlength{\parskip}{-10pt}
        \begin{subfigure}{0.48\linewidth}
                \subcaption{}
                \includegraphics[width=\linewidth]{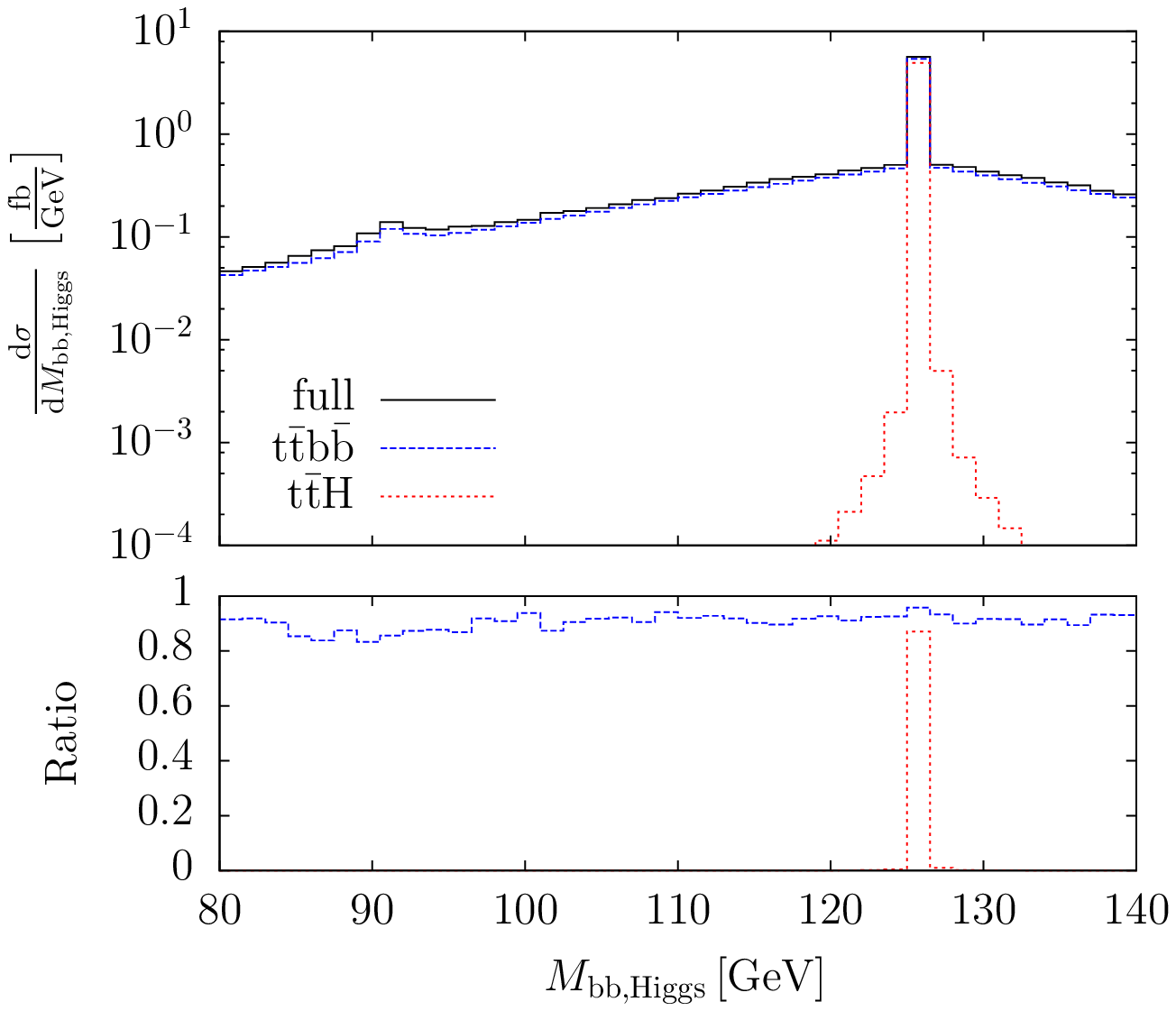}
                \label{plot:invariant_mass_best_higgs_mass_bjet_pair}%
        \end{subfigure}
        \hfill
        \begin{subfigure}{0.48\linewidth}
                \subcaption{}
                \includegraphics[width=\linewidth]{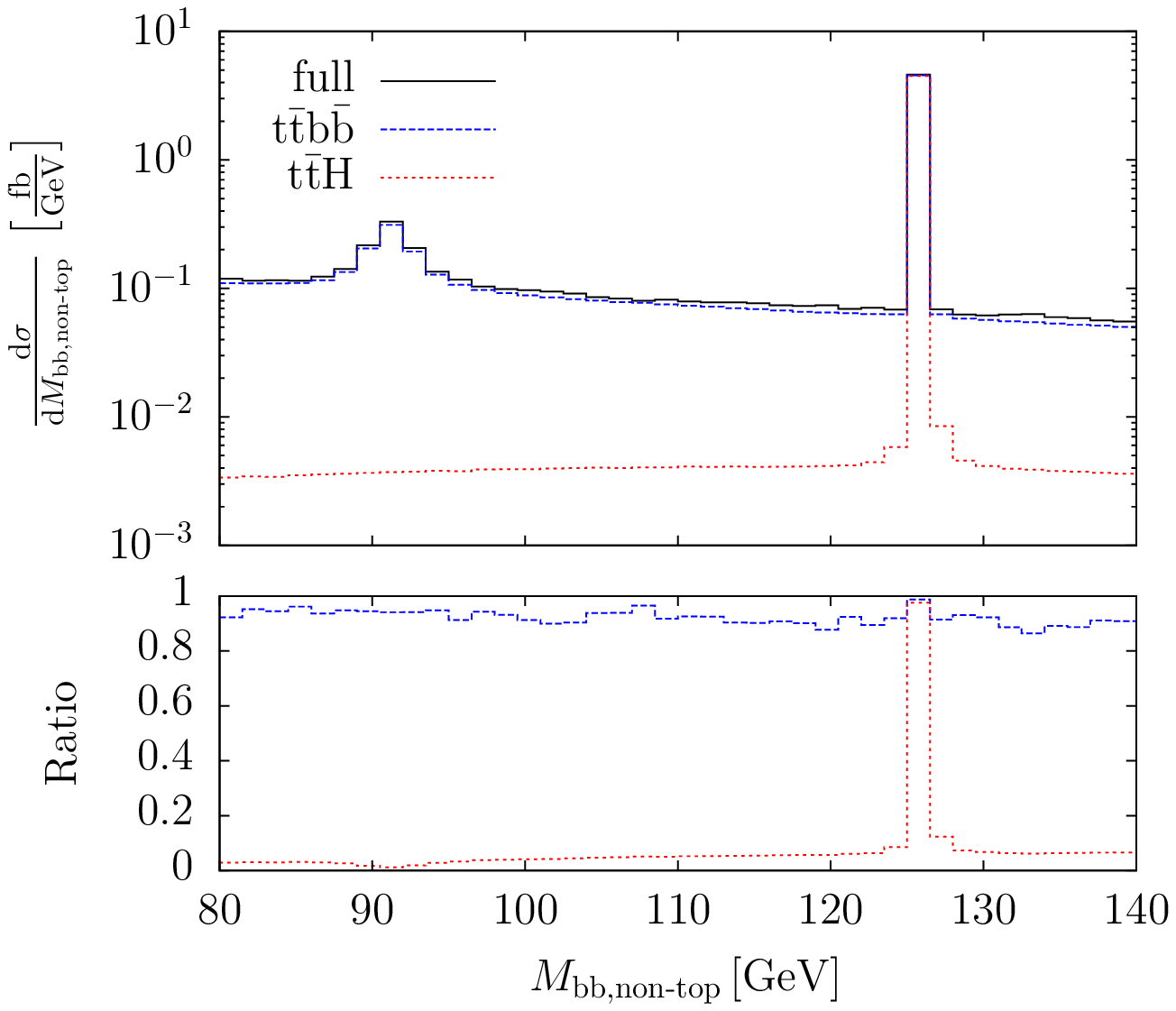}
                \label{plot:invariant_mass_min_llh_bjet_pair}%
        \end{subfigure}
        
        \begin{subfigure}{0.48\linewidth}
                \subcaption{}
                \includegraphics[width=\linewidth]{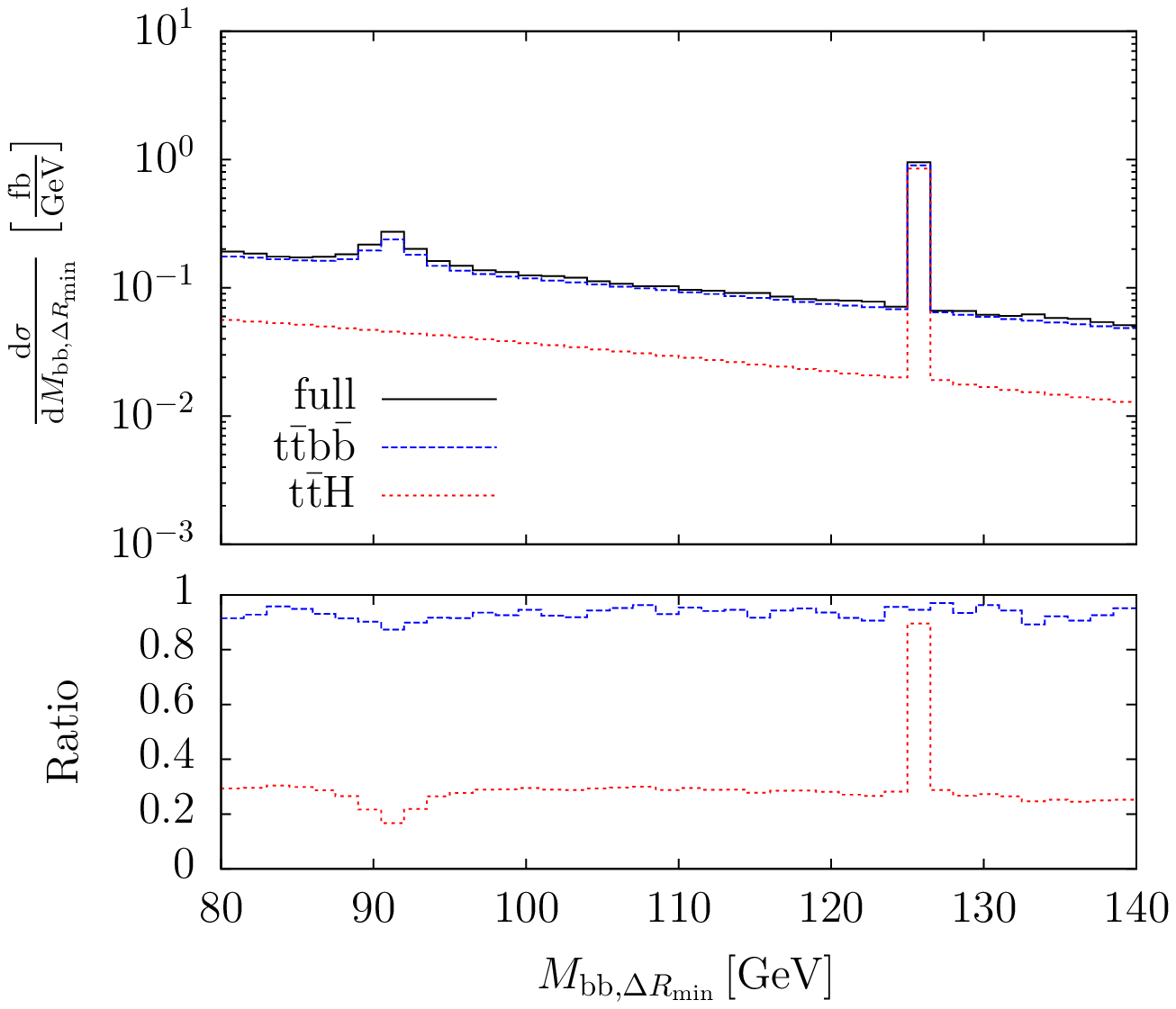}
                \label{plot:invariant_mass_minimal_distance_bjet_pair}%
        \end{subfigure}
        \hfill
        \begin{subfigure}{0.48\linewidth}
                \subcaption{}
                \includegraphics[width=\linewidth]{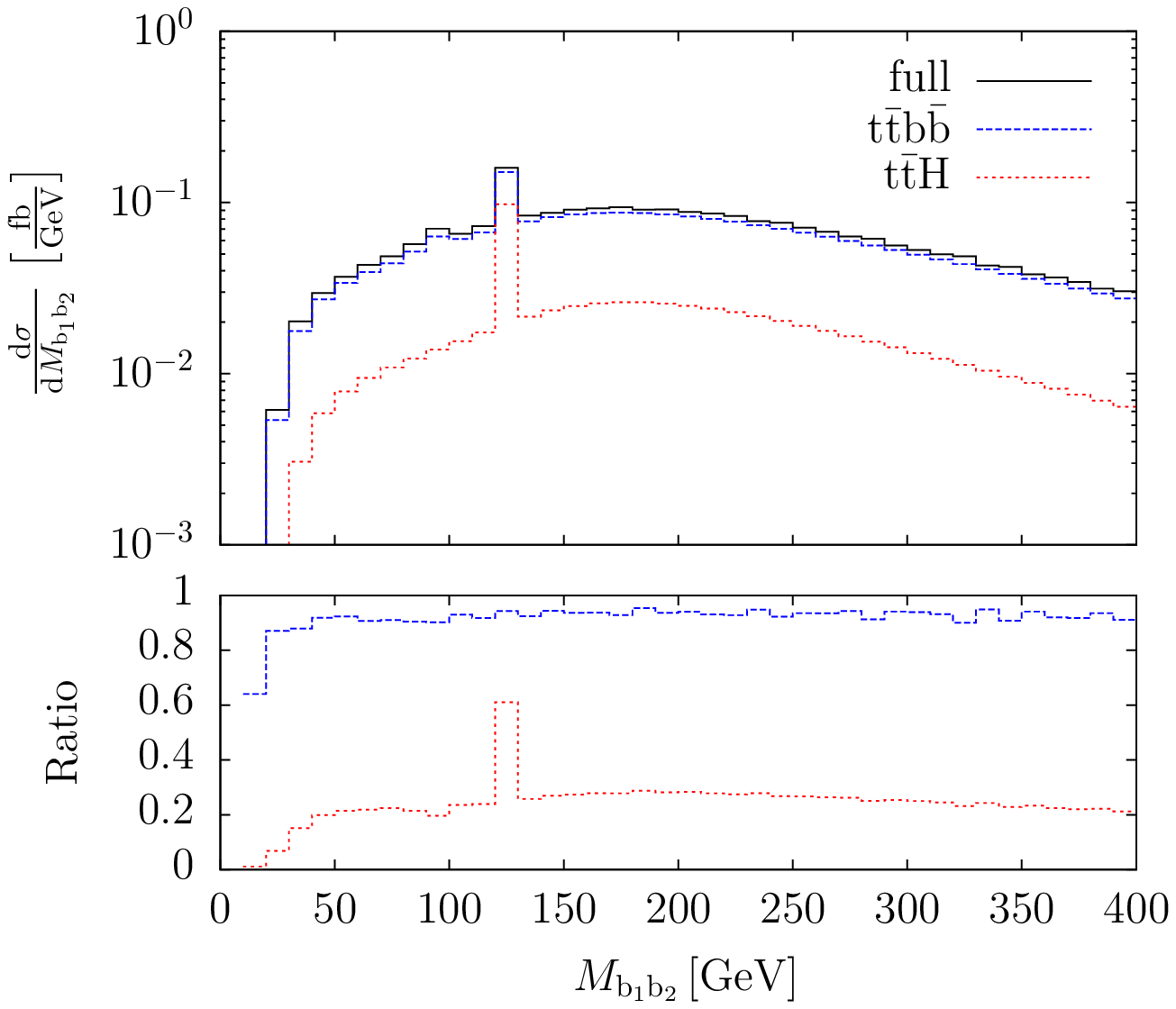}
                \label{plot:invariant_mass_hardest_bjet_pair}%
        \end{subfigure}%

        \caption{\label{fig:Mbb_one}
                Invariant-mass distributions at the LHC at $13\TeV$ of the b-jet pair that %
          \subref{plot:invariant_mass_best_higgs_mass_bjet_pair} forms an invariant mass closest to the Higgs-boson mass (upper left), %
          \subref{plot:invariant_mass_min_llh_bjet_pair} forms an invariant mass determined by top--antitop Breit--Wigner maximum likelihood (upper right), %
          \subref{plot:invariant_mass_minimal_distance_bjet_pair} forms the smallest distance $\Delta R$ (lower left), %
          \subref{plot:invariant_mass_hardest_bjet_pair} has the highest transverse momenta (lower right). %
          The lower panels show the relative size of $\ttbbb$ and
          $\ttbarh$ production normalised to the full process.}
\end{figure}
For this observable we compute the
invariant masses of all six b-jet pairs and choose the one that is
closest to the Higgs-boson mass used in the Monte Carlo run.
Figure~\ref{plot:invariant_mass_best_higgs_mass_bjet_pair} displays
the resulting invariant-mass distribution.  For \ttbarh production it
features a very clear peak at the Higgs-boson mass with a strong
drop-off according to the Breit--Wigner shape of the Higgs-boson
resonance. Thus, the ratio of \ttbarh production and the full process
almost vanishes outside the resonant region. The finite off-shell
contributions are a result of the pole approximation where the
on-shell projected momenta are only applied to the matrix element
evaluation, but not to the resonant propagator. The phase-space
integration and thus the histogram binning is performed with off-shell
momenta.  The \ttbarbbbar and the full process have contributions
where the Higgs boson is replaced by a Z~boson, visible as a bump
around the Z-boson mass in the differential distribution.

This observable is strongly biased, since we always choose the b-jet
combination that gives the invariant mass closet to the Higgs-boson
mass. This explains the rise of the differential cross section for
$\ttbbb$~production and the full process towards the Higgs resonance
outside the peak.

\subsubsection{Invariant mass of b-jet pair determined by top--antitop 
Breit--Wigner maximum likelihood}
\label{sec:likelihood-mass}

Motivated by \citere{ATLAS:2012cpa} we determine the two b jets that
most likely originate from the decay of the top quark ($\Pt\to\PW^+
\Pb\to\Pl^+\nu_\Pl\Pb$) and antitop quark
($\bar{\Pt}\to\PW^-\bar{\Pb}\to\bar{u}d\bar{\Pb}$, with $u=\Pu, \Pc$,
$d=\Pd, \Ps$) and plot the invariant mass of the remaining b-jet pair.
Since in most events the top quark and antitop quark in \ttbarh
production are nearly on-shell, the two b jets maximising the
corresponding propagator contributions are most likely to originate
from the top-quark and antitop-quark decay. To determine the
maximising b-jet combination we compute a top-momentum candidate with
the charged lepton, neutrino and a b-jet momentum ($p_{\text{b}_i}$),
\begin{equation}
        p_{\Pl^+\nu_\Pl \text{b}_i} = p_{\Pl^+} + p_{\nu_\Pl} + p_{\Pb_i}
\end{equation}
and an analogous antitop-momentum candidate with the two momenta of
the non-b jets and a 
different b-jet momentum ($p_{\Pb_j}$),
\begin{equation}
        p_{\text{j}_1\text{j}_2\text{b}_j} = p_{\text{j}_1} + p_{\text{j}_2} + p_{\text{b}_j}.
\end{equation}
As b jets originating from the top-quark and antitop-quark decay we
select those that maximise the likelihood function $\mathcal{L}$
defined as a product of two Breit--Wigner distributions corresponding
to the top-quark and antitop-quark propagators:
\begin{equation}
        \mathcal{L} \propto \frac{1}{\Bigl(p_{\Pl^+\nu_\Pl \Pb_i}^2 \!- \Mt^2\Bigr)^2+\left(\Mt\Gt\right)^2}\, \frac{1}{\Bigl(p_{\text{j}_1\text{j}_2\text{b}_j}^2\!- \Mt^2\Bigr)^2+\left(\Mt\Gt\right)^2}.
\end{equation}

In \reffi{plot:invariant_mass_min_llh_bjet_pair} we present the
b-jet-pair invariant mass that has been identified to originate from
the Higgs-boson decay by the maximum-likelihood method described
above.  In the off-shell region the ratio of \ttbarh production to the
full process drops considerably below the corresponding ratio for the
total cross section of about a fourth. Owing to the absence of the
bias, the peak in the resonant region is more pronounced for the full
process compared the method based on the b-jet-pair invariant mass
closest to the Higgs-boson mass
(\reffi{plot:invariant_mass_best_higgs_mass_bjet_pair}), yielding
a better signal to background ratio.  Since this method tags the
b~jets resulting from the top and antitop quarks, any resonance in the
invariant mass of the remaining b-jet pair is resolved, and thus the
Z~resonance is clearly visible in the plot.

\subsubsection[Invariant mass of minimal
\texorpdfstring{\deltar}{Delta R}-distance b-jet pair]
{\boldmath Invariant mass of minimal \deltar-distance b-jet pair}
\label{subsubsec:Minimal_distance_Method}

Next we compute for each event the \deltar distance of all six b-jet
pairs and plot the invariant mass of the pair with the smallest
\deltar distance.  Since the Higgs boson, as a scalar particle, decays
isotropically, this observable is not sensitive to Higgs-boson
production at the $\ttbarh$ threshold but potentially for boosted
Higgs bosons.
Figure~\ref{plot:invariant_mass_minimal_distance_bjet_pair} displays
the corresponding invariant-mass distribution for all three scenarios.
The Higgs-boson peak is clearly visible in all cases.  However, it is
less pronounced compared to the Breit--Wigner maximum-likelihood
method (Figure~\ref{plot:invariant_mass_min_llh_bjet_pair}) and the
ratio of peak over background is only weakly enhanced compared to the
purely combinatorial effect.  Outside the regions of the Higgs and
Z-boson resonances the ratio of \ttbarh production vs.\ the full
process equals the average value of about a fourth, which we see in
the total cross-section ratio.  This indicates that the b-jet pair
with minimal \deltar~distance is almost evenly distributed among all
six possible b-jet pairs, i.e.\ the method fails to tag the b-jet pair
originating from the Higgs-boson decay.

The bump around the Z-boson mass in the differential distributions
is weaker as compared to \reffi{plot:invariant_mass_min_llh_bjet_pair}
and there appears a ditch in the \ttbarh ratio because the Z-boson resonance
is absent in \ttbarh production. The ditch in the \ttbarbbbar ratio
is due to relative differences in contributions from resonant \PZ
bosons in the full process and in  \ttbarbbbar production. 
Thus, in the considered set-up this method is not suitable to tag the 
Higgs boson. This might be different if strongly boosted Higgs bosons
are required. 

\subsubsection{Invariant mass of two hardest b~jets}
For comparison we show the distribution in the invariant mass of the
pair of the two hardest b~jets in
\reffi{plot:invariant_mass_hardest_bjet_pair}. Also here the
Higgs-boson signal is clearly visible for all processes but there is
no enhancement above the purely statistical level. The apparently
smaller enhancement of the Higgs peak as compared to
\reffi{plot:invariant_mass_minimal_distance_bjet_pair} is basically
due to the larger bin width ($10\GeV$ as compared to $1.5\GeV$).
Furthermore, outside the resonance regions \ttbarh production is about
a fourth of the full process, as for the total cross section.

\subsubsection{Further differential distributions}
In this section we investigate further differential distributions
concentrating on observables that show deviations in the shape
between the full process and the approximations. Some other
distributions which show no significant shape
deviations are listed in Appendix~\ref{sec:FurtherObs}.
\begin{figure}
        \setlength{\parskip}{-10pt}
        \begin{subfigure}{0.48\linewidth}
                \subcaption{}
                \includegraphics[width=\linewidth]{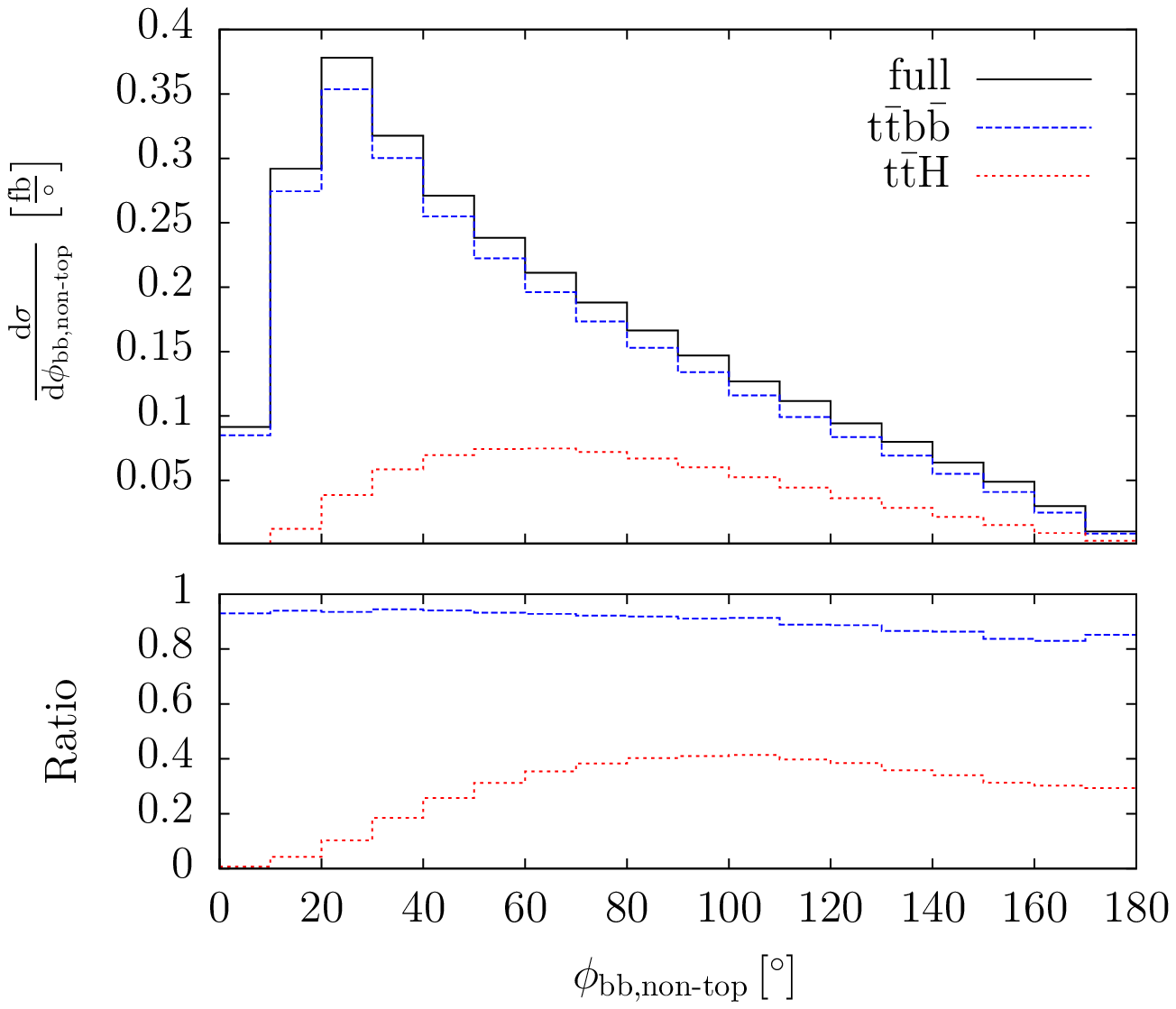}
                \label{plot:azimuth_separation_min_llh_bjet_pair}
        \end{subfigure}
        \hfill
        \begin{subfigure}{0.48\linewidth}
                \subcaption{}
                \includegraphics[width=\linewidth]{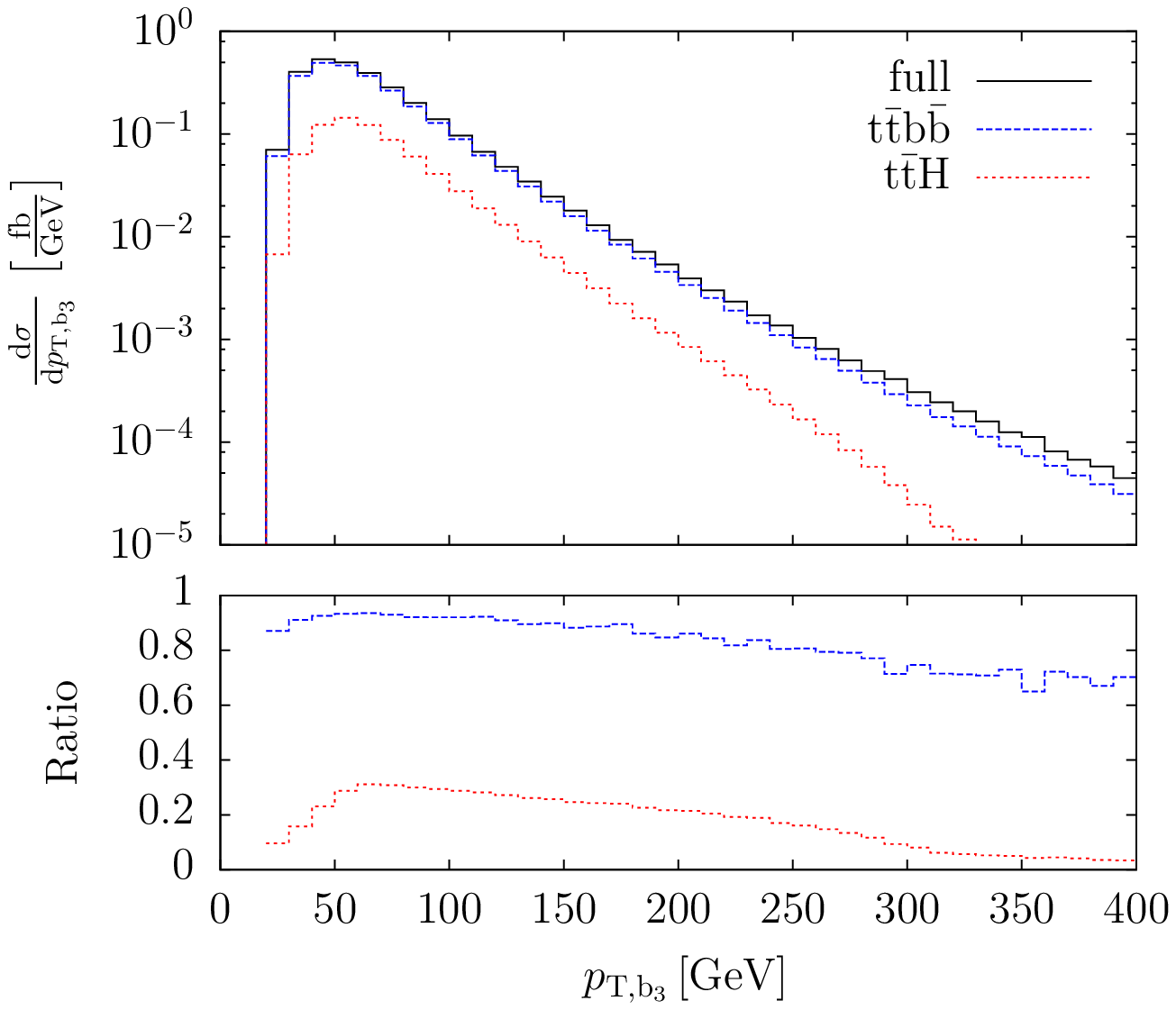}
                \label{plot:transverse_momentum_3rd_hardest_bjet} 
        \end{subfigure}
        
        \begin{subfigure}{0.48\linewidth}
                \subcaption{}
                \includegraphics[width=\linewidth]{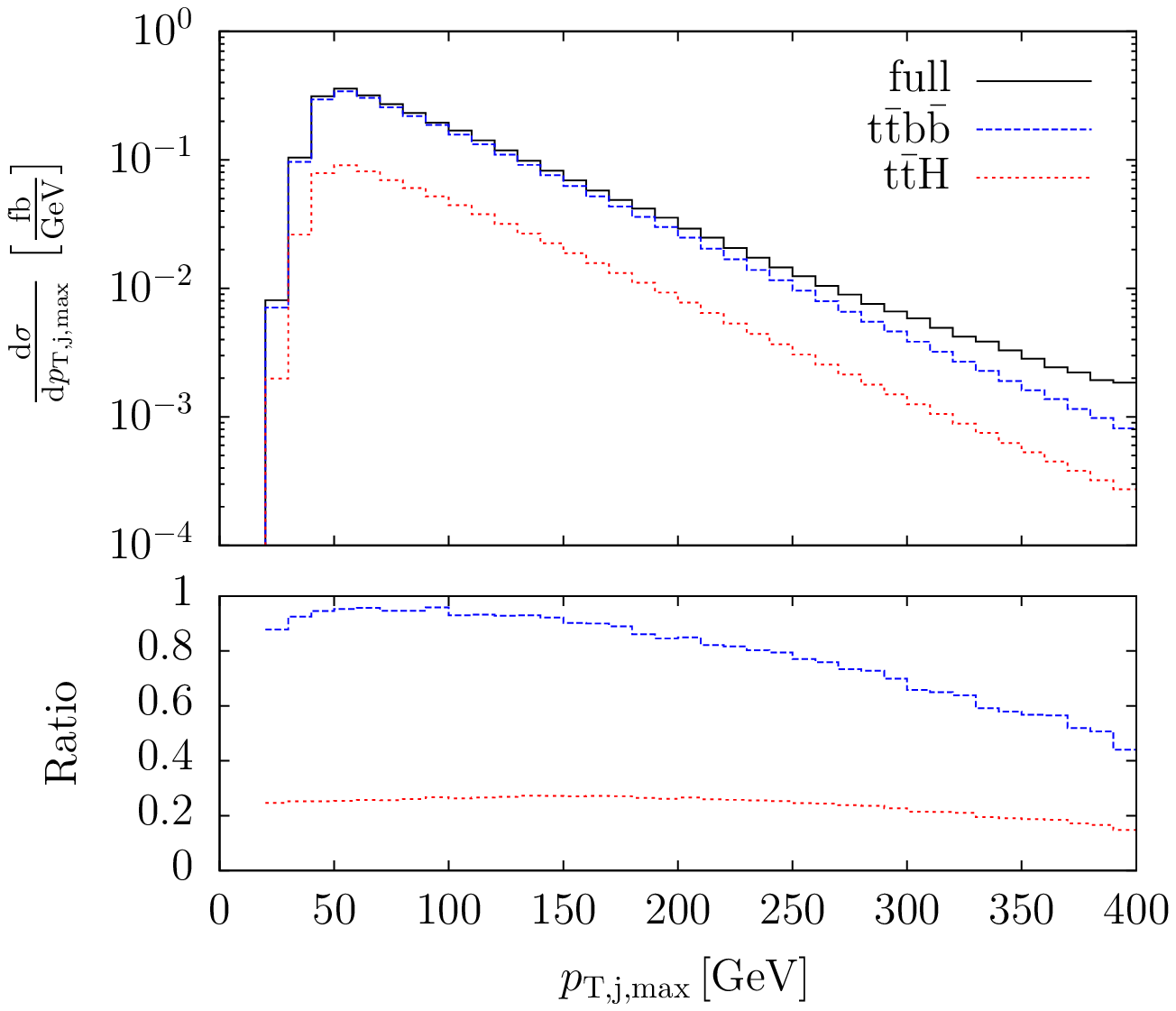}
                \label{plot:transverse_momentum_hardest_jet}
        \end{subfigure}
        \hfill
        \begin{subfigure}{0.48\linewidth}
                \subcaption{}
                \includegraphics[width=\linewidth]{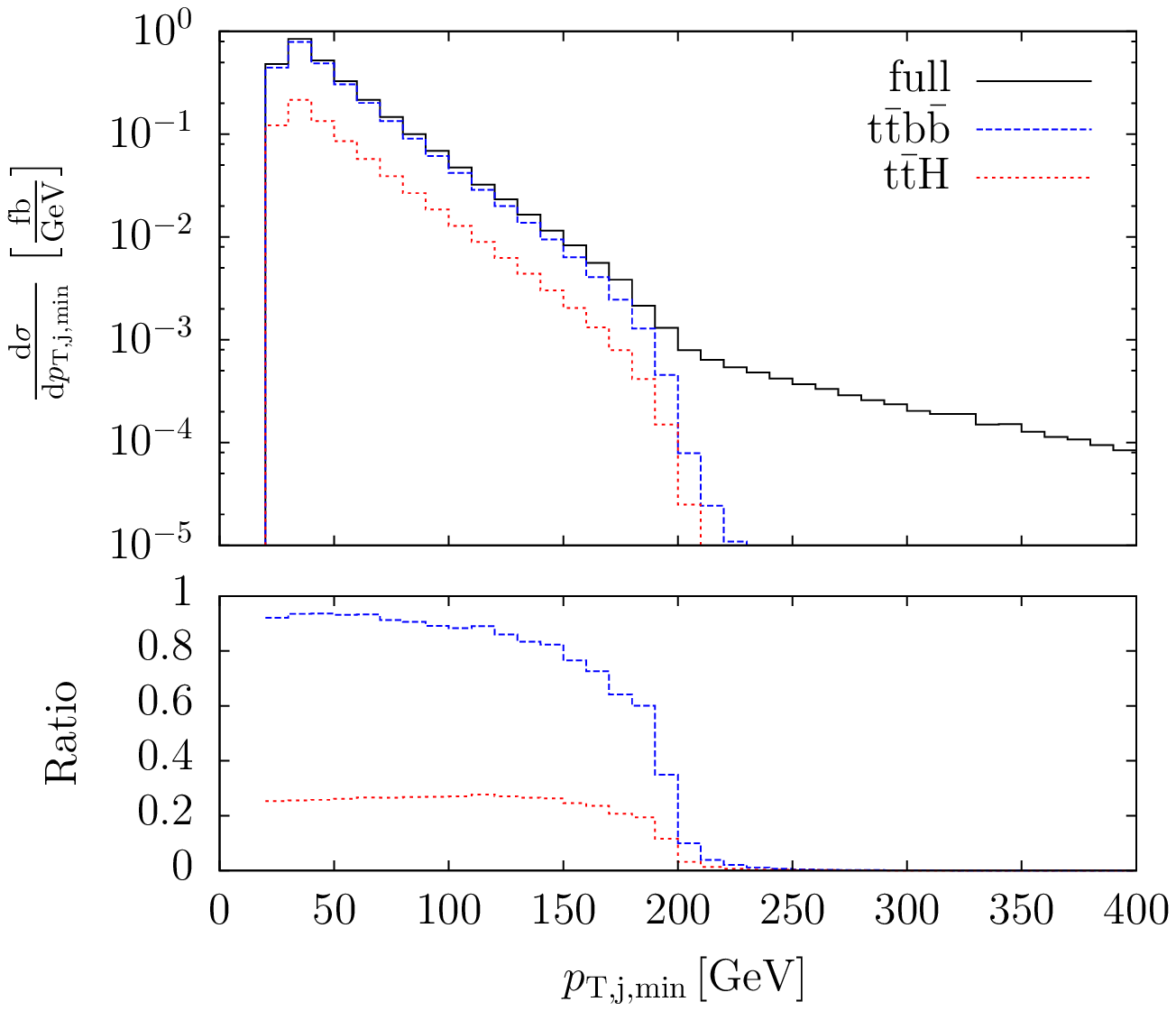}
                \label{plot:transverse_momentum_softest_jet}
        \end{subfigure}
        
        \begin{subfigure}{0.48\linewidth}
                \subcaption{}
                \includegraphics[width=\linewidth]{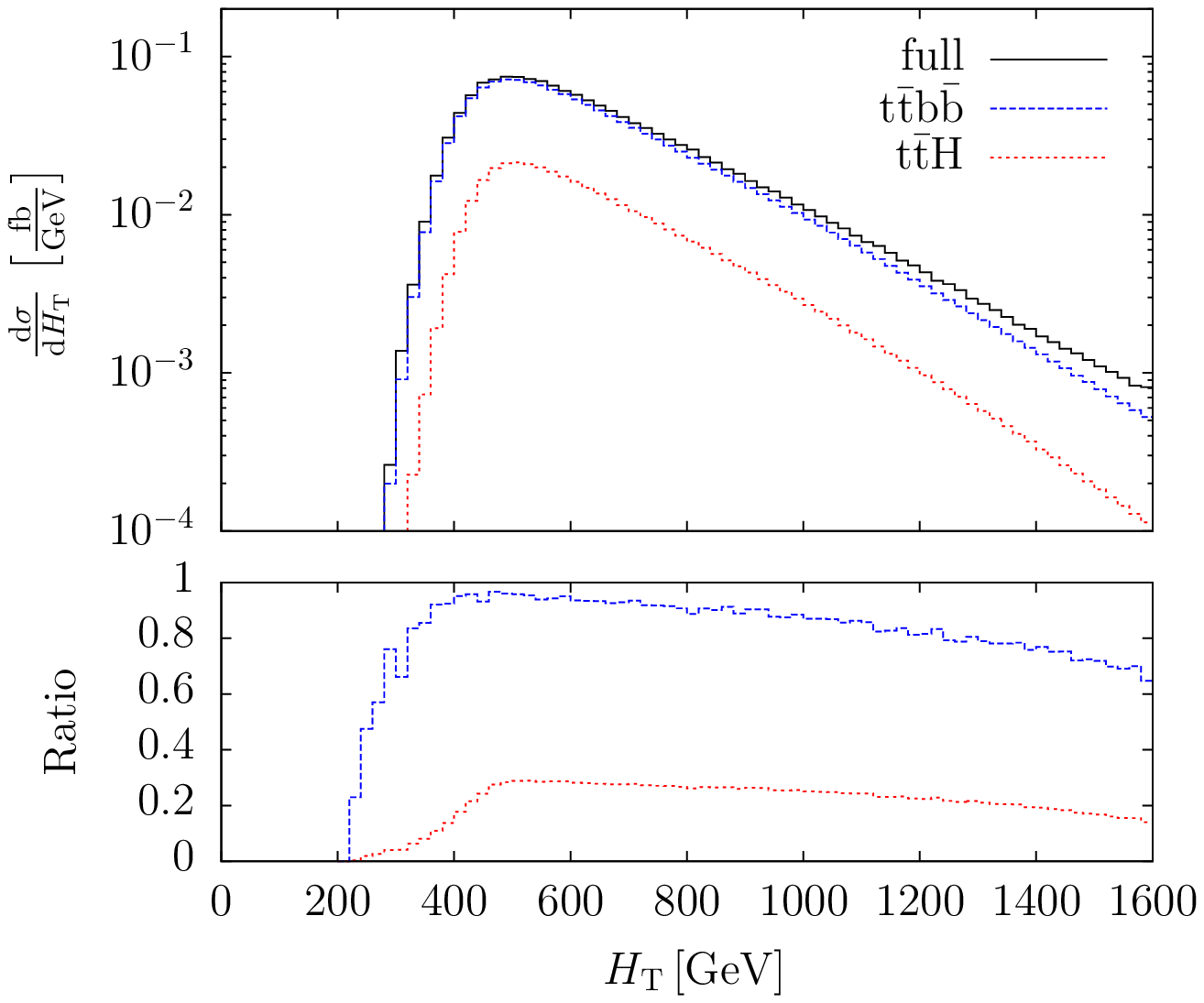}
                \label{plot:total_transverse_energy_born_lo}
        \end{subfigure}
        \hfill
        \begin{subfigure}{0.48\linewidth}
                \subcaption{}
                \includegraphics[width=\linewidth]{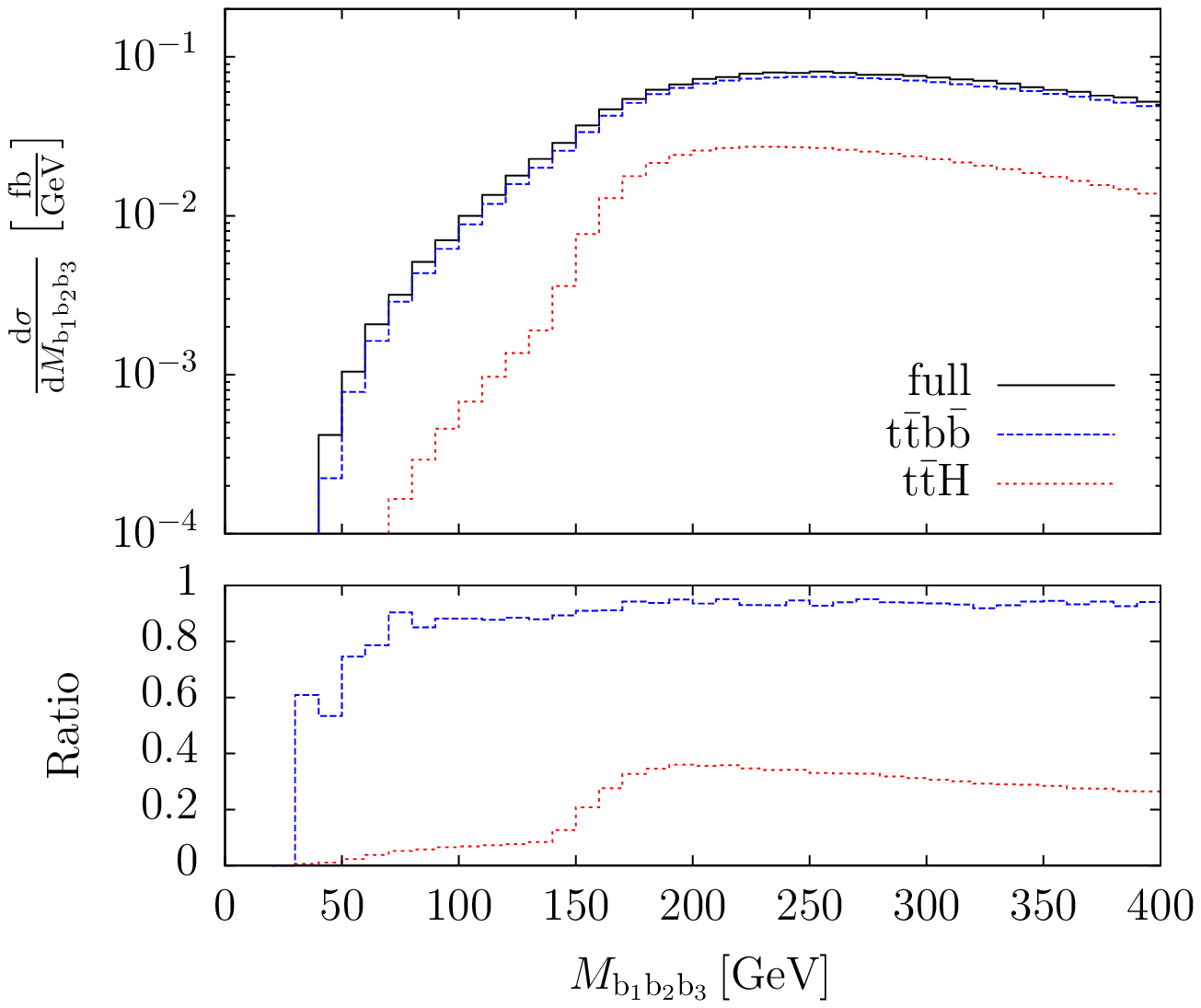}
                \label{plot:invariant_mass_bbb123_hardest_bjet}
        \end{subfigure}%
        \vspace*{-3ex}
        \caption{%
                Differential distributions at the LHC at $13\TeV$ exhibiting shape deviations between the full process and
                \ttbarbbbar or \ttbarh production:     
                \subref{plot:azimuth_separation_min_llh_bjet_pair}~azimuthal
                separation of the b-jet pair determined by top--antitop
                Breit--Wigner maximum likelihood (upper left), %
                \subref{plot:transverse_momentum_3rd_hardest_bjet}~transverse momentum of the 3rd-hardest \Pb~jet (upper right), %
                \subref{plot:transverse_momentum_hardest_jet} transverse momentum of the hardest non-\Pb~jet (middle left), %
                \subref{plot:transverse_momentum_softest_jet} transverse momentum of the softest non-\Pb~jet (middle right), %
                \subref{plot:total_transverse_energy_born_lo} total transverse energy (lower left), %
                \subref{plot:invariant_mass_bbb123_hardest_bjet} invariant mass of the three hardest \Pb~jets (lower right). %
                The lower panels show the relative size of $\ttbbb$ and $\ttbarh$ production normalised to the full process.}
\end{figure}%

Figure~\ref{plot:azimuth_separation_min_llh_bjet_pair} shows the
azimuthal separation of the b-jet pair determined by top--antitop
Breit--Wigner maximum likelihood according to
\refse{sec:likelihood-mass}. While \ttbarbbbar production and the full
process yield a very similar shape, \ttbarh production exhibits
clearly a different shape.  This behaviour can be explained by the
dominant production mechanisms of bottom--antibottom pairs. In the
signal process these result from the Higgs boson and owing to the
finite Higgs-boson mass tend to have a finite opening angle. In the
background processes the bottom--antibottom pairs result mainly from
gluons and thus tend to be collinear leading to a peak at small
$\phi_{\Pb\Pb}$ that is cut off by the acceptance function.  Thus,
this distribution can help to separate bottom--antibottom pairs
resulting from Higgs bosons from those of other origin.

Figure~\ref{plot:transverse_momentum_3rd_hardest_bjet} displays the
transverse-momentum distribution of the third-hardest b jet. We find
that all three approximations are similar in shape for
$p_\text{T}$~values below $150\GeV$. For higher transverse momenta the
distribution for \ttbarh production diverges from those of the full
process and \ttbarbbbar production. We do not see this behaviour in
the transverse momentum distributions of the two harder b~jets (see
\reffi{plot:transverse_momentum_hardest_bjet} in
Appendix~\ref{sec:FurtherObs}) but to some extent in the one of the
fourth-hardest b jet. This results from the fact that in the $\ttbarh$
signal all b~jets originate from heavy-particle decays, while in the
full process some are directly produced yielding more b~jets with high
transverse momenta.

The distributions in the transverse momenta of the two non-b jets are
displayed in
\reffis{plot:transverse_momentum_hardest_jet}--\ref{plot:transverse_momentum_softest_jet}.
For the hardest non-b jet, we find a similar picture as for the 3rd
hardest b jet and an enhancement for the full process relative to
\ttbarbbbar production and \ttbarh production for transverse momenta
above $150\GeV$. The explanation is similar as in the preceding case.
While in the \ttbarbbbar process the jets originate from the top-quark
decay, in the full process they can be produced directly leading to
more jet activity at high transverse momenta.  In the case of the
second hardest jet both approximations exhibit a strong drop near
{$p_{\rm T,j,min}=200\GeV$}. For higher transverse momenta two jets
originating from W-boson decay are too collinear to pass the
rapidity--azimuthal-angle-separation cut of $\Delta R_{\Pj\Pj}>0.4$ such that
the corresponding events are eliminated. On the other hand, events with
jets pairs with higher invariant masses, which are present in the full
process, are not cut.

The sum of all transverse energies (including missing transverse
energy) is depicted in \reffi{plot:total_transverse_energy_born_lo}. For
small $H_{\rm T}$ the different thresholds of the approximations are
clearly visible. For $H_{\rm T}\sim 400\GeV$--$800\GeV$ the $\ttbbb$
approximation describes the full process within $10\,\%$. As it
decreases stronger with increasing $H_{\rm T}$ the deviation becomes
larger above $800\GeV$.  Since $H_{\rm T}$ incorporates all transverse
energies of the process it is a measure for the average deviation of
the transverse energies between the approximations and the full
process.

Finally, \reffi{plot:invariant_mass_bbb123_hardest_bjet} presents the
invariant mass of the three hardest b jets. Below the threshold
$\MH+p_{\mathrm{T},\Pb,\mathrm{cut}}\approx150\GeV$ the signal process
is strongly suppressed, above its ratio to the full process rises to
$36\,\%$ at $M_{\mathrm{b_1b_2b_3}}\sim 195\GeV$ and then drops slowly
to $26\,\%$ at $M_{\mathrm{b_1b_2b_3}}\sim 400\GeV$. The ratio of
\ttbarbbbar production and the full process on the other hand is
roughly constant for $M_{\mathrm{b_1b_2b_3}}\gsim 70\GeV$.

\subsection{Interference effects in differential distributions}
\label{sec:interference_distributions}  
                
\begin{figure}
        \setlength{\parskip}{-5pt}
        \captionsetup[subfigure]{skip=2pt}
        \begin{subfigure}{0.48\linewidth}
                \subcaption{}
                \label{plot:interference_transverse_momentum_third_hardest_bjet}
                \includegraphics[width=\linewidth]{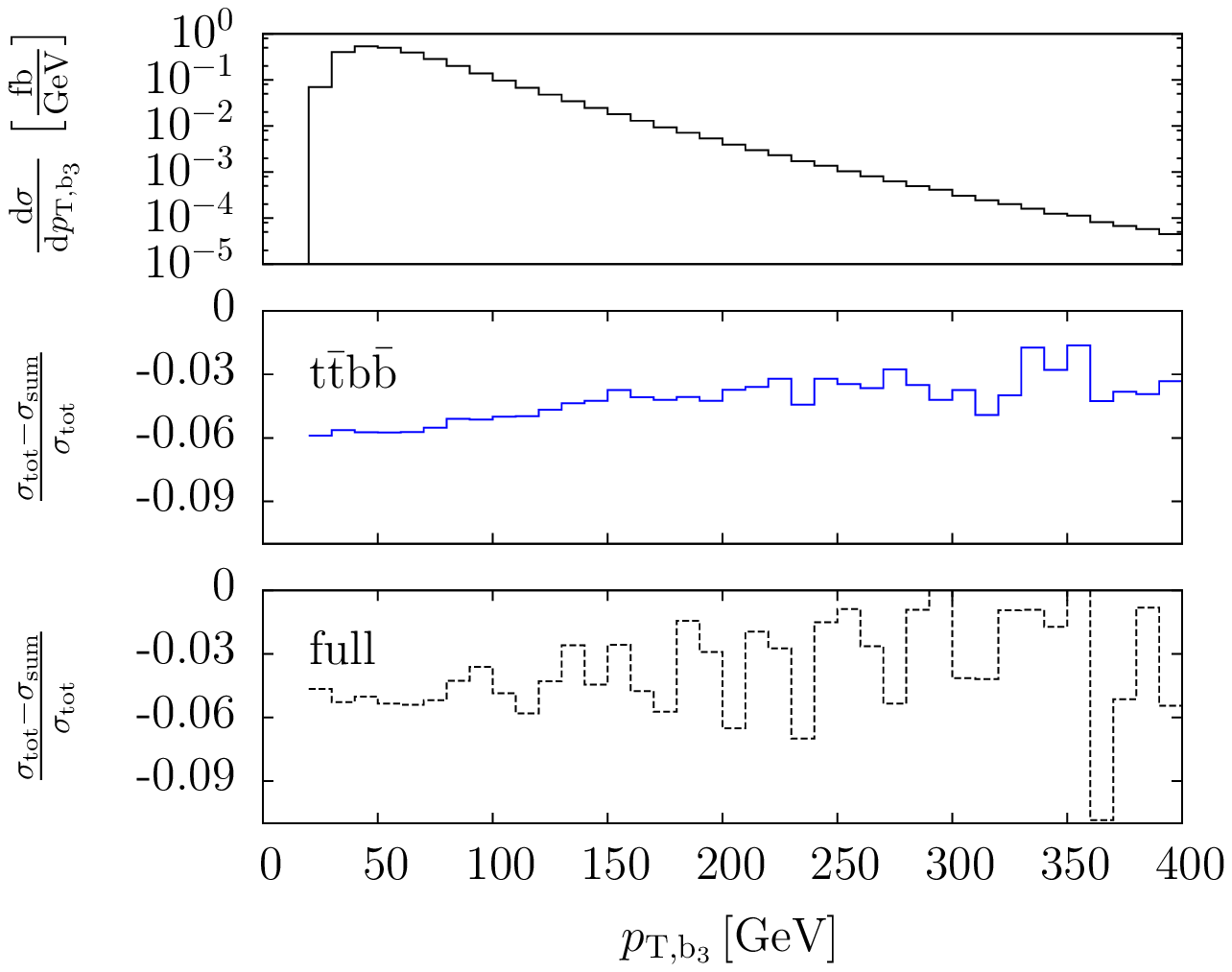}
        \end{subfigure}
        \hfill
        \begin{subfigure}{0.48\linewidth}
                \subcaption{}
                \label{plot:interference_pt_jet_harder}
                \includegraphics[width=\linewidth]{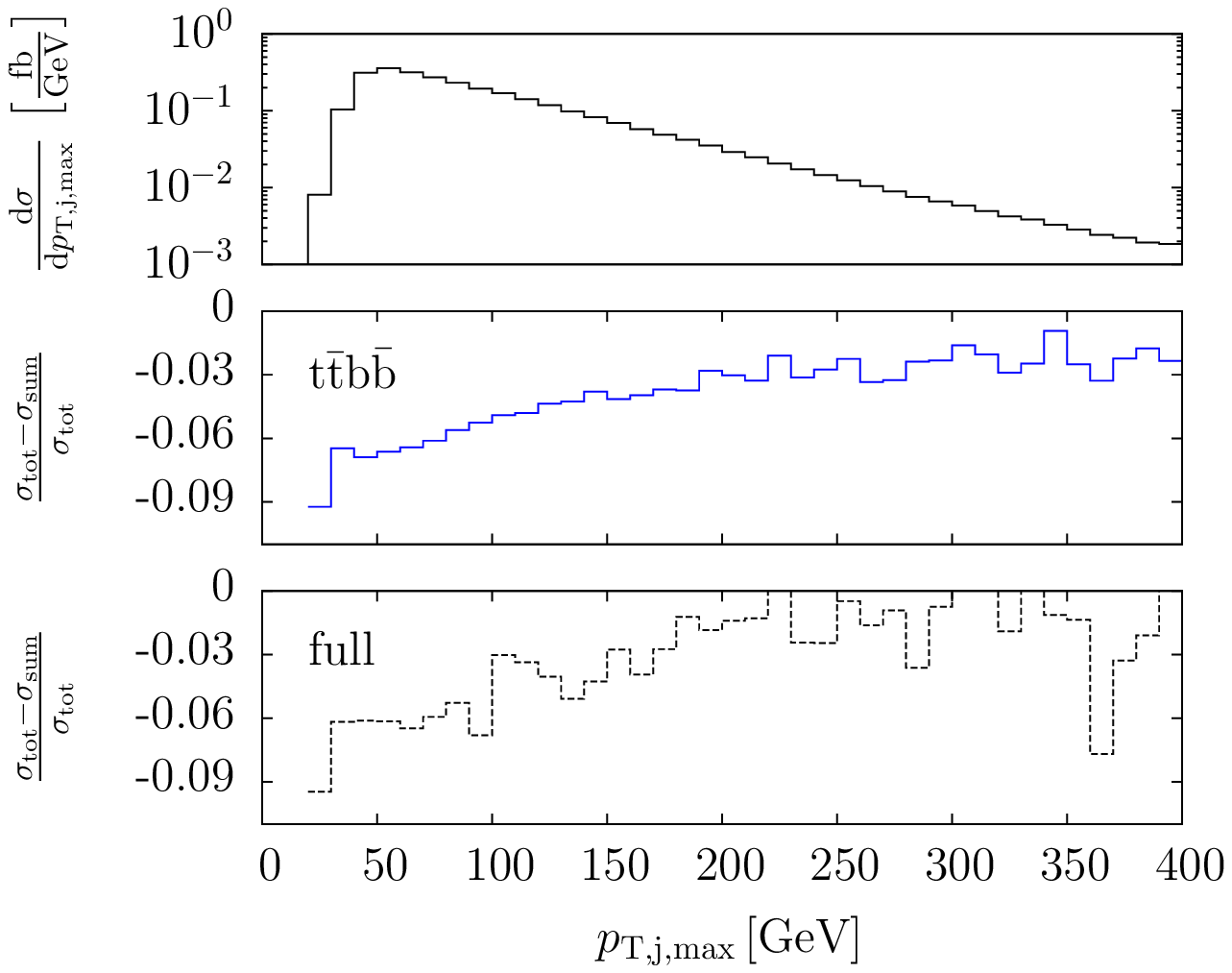}
        \end{subfigure}
        
        \begin{subfigure}{0.48\linewidth}
                \subcaption{}
                \label{plot:interference_invariant_mass_min_llh_bjet_pair}
                \includegraphics[width=\linewidth]{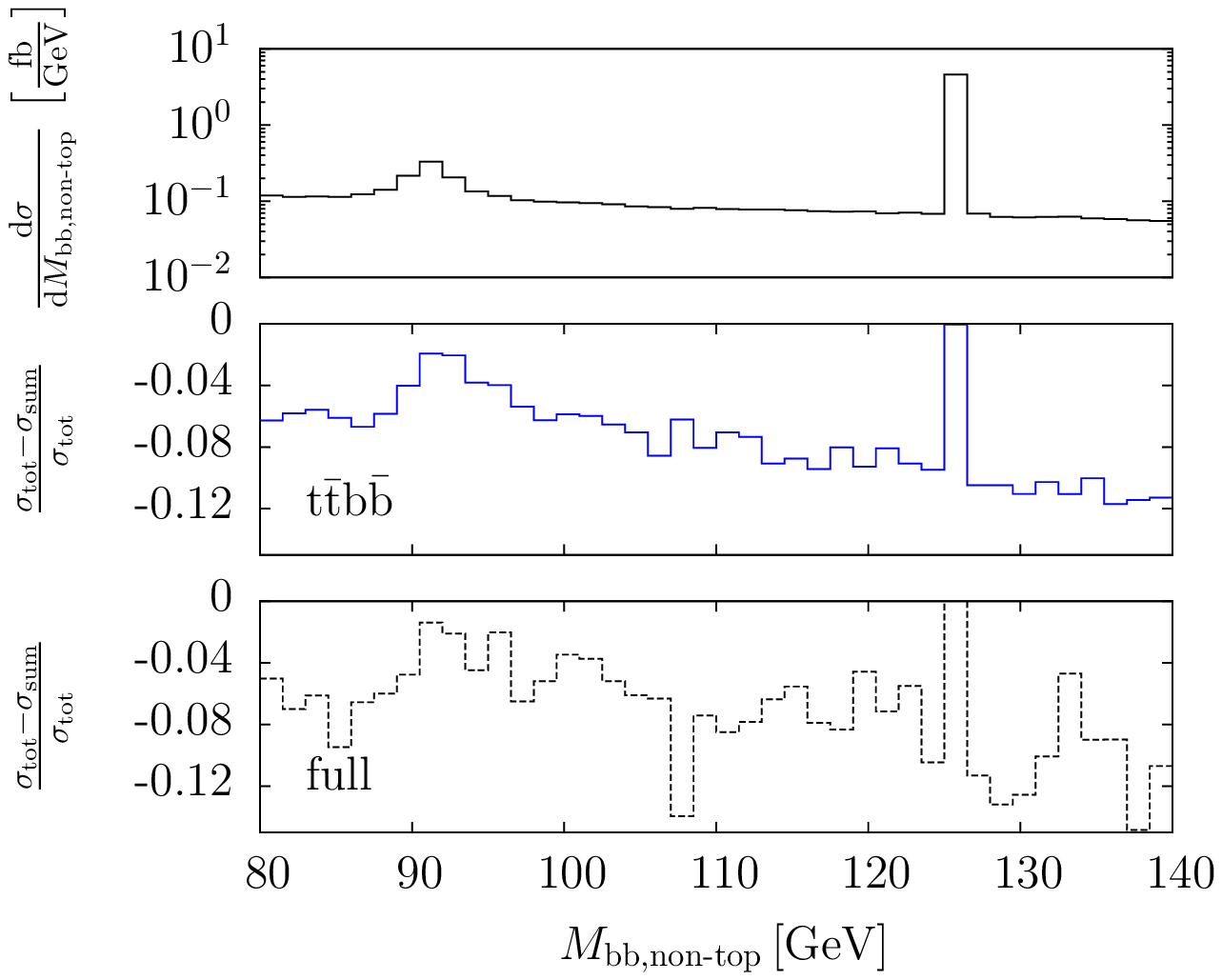}
        \end{subfigure}
        \hfill
        \begin{subfigure}{0.48\linewidth}
                \subcaption{}
                \label{plot:interference_azimuth_separation_min_llh_bjet_pair} 
                \includegraphics[width=\linewidth]{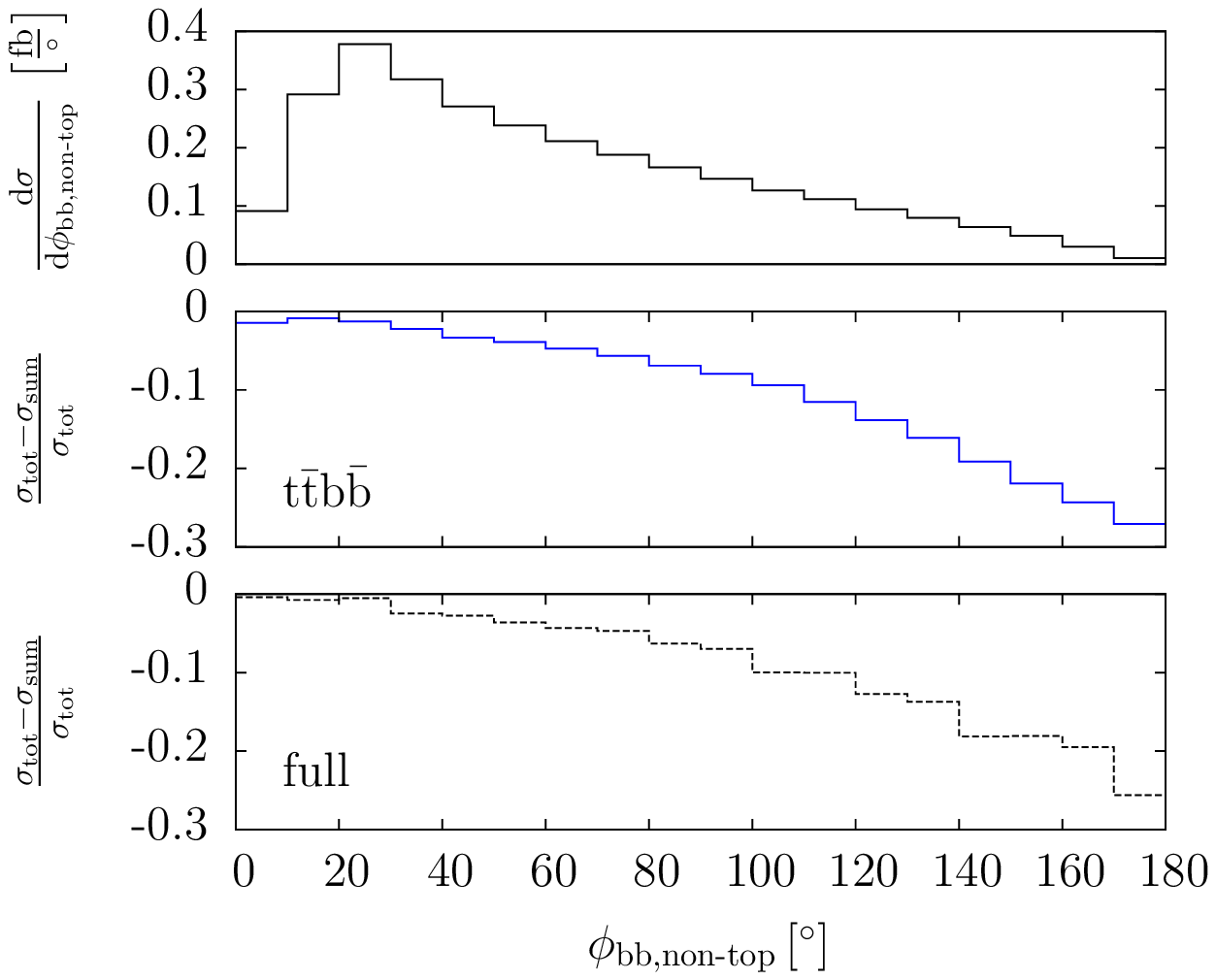}
        \end{subfigure}
        \caption{\label{plot:interference}
                Interference effects versus     
                \subref{plot:interference_transverse_momentum_third_hardest_bjet} transverse momentum of the 3rd-hardest \Pb~jet (upper left), %
                \subref{plot:interference_pt_jet_harder} transverse momentum of the hardest non-\Pb~jet (upper right), %
                \subref{plot:interference_invariant_mass_min_llh_bjet_pair} invariant mass of the of the b-jet pair determined by top--antitop Breit--Wigner maximum likelihood (lower left), %
                \subref{plot:interference_azimuth_separation_min_llh_bjet_pair} azimuthal separation of the b-jet pair determined by top--antitop Breit--Wigner maximum likelihood (lower right). %
                The lower panels show the relative interference effects of
                $\ttbbb$ production and the full process, respectively.  The
                upper panel shows the corresponding differential
                distribution of the full process as reference.}
\end{figure}
In this section we study in detail the effects of the interference
contributions between matrix elements of different orders in \alphas.
For most distributions we find a uniform shift by roughly the same
amount as for the total cross section, i.e.\ about $5\,\%$ for
\ttbarbbbar production and the full process (interference effects are
absent in \ttbarh production). For both scenarios we observe a few
kinematical distributions that are sensitive to these interference
effects.  The upper panels of \reffi{plot:interference} show the
results for the full process and the central and lower panels
highlight the interference effects.  Specifically, the central panels
show the relative difference
${(\sigma_\text{tot}-\sigma_\text{sum})}/{\sigma_\text{tot}}$ for
\ttbarbbbar production with a solid blue line and the lower panels the
same relative difference for the full process with a dashed line.

In  
\reffis{plot:interference_transverse_momentum_third_hardest_bjet} and
\ref{plot:interference_pt_jet_harder} we see that the effect of the
interference for the distributions in the transverse momentum of the
third-hardest b jet and for the harder non-b jet, respectively, varies
monotonically with increasing 
$p_\text{T}$ from $-6\,\%$ to $-2\,\%$.
Figure~\ref{plot:interference_invariant_mass_min_llh_bjet_pair} shows 
the interference effects on the distribution of the invariant mass of the b-jet pair 
determined by top--antitop Breit--Wigner maximum likelihood. The
suppression of interference in the regions of the Higgs- and Z-boson
resonances is clearly visible. For invariant masses above the Higgs threshold the
interference effect exceeds $-10\,\%$.
As shown in
\reffi{plot:interference_azimuth_separation_min_llh_bjet_pair}, the
relative interference effects grow with increasing azimuthal-angle
separation of the b-jet pair determined by top--antitop Breit--Wigner
maximum likelihood from almost zero at small angles to $-25\,\%$ for
$\phi_{\mathrm{bb,non\textnormal{-}top}}=180^\circ$, while the cross section
drops with increasing azimuthal-angle separation.  Also in the
distributions the dominant interference effects arise from diagrams of
order \order{\alphas\alpha^3} involving $t$-channel W-boson exchange
interfering with diagrams of order \order{\alphas^2\alpha^2}.
Interferences with diagrams for $\ttbarh$ production are by a factor
five or more smaller.

\section{Conclusion and outlook}
\label{sec:conclusion}

We have presented an analysis of the irreducible background for
$\ttbarh$ production at the LHC.  Specifically, we have compared the
full Standard Model cross section for the production of four b jets,
two jets, one identified charged lepton and missing energy with the
contributions from the subprocesses \ttbarbbbar production and \ttbarh
production, obtained using the pole approximation. With standard
acceptance cuts we find that the total cross section of \ttbarh
production and decay is roughly a fourth of the full process, while
\ttbarbbbar production constitutes the major contribution to the full
process with about $92\,\%$. For all scenarios the bulk of the cross
section originates from gluon-induced processes.

We analysed various b-jet-pair invariant-mass distributions based on
different methods to select two of the four b jets to be identified
with the decay products of the Higgs boson.  We find that assigning
two b jets to the top- and antitop-quark decay by maximising a
combined Breit--Wigner likelihood function and assigning the remaining
two b jets to the potential Higgs boson yields a good unbiased
determination of the b-jet pair originating from the Higgs-boson
decay.

We investigated the interferences between contributions to the matrix
element of different orders in the strong and electroweak couplings
constants.  We find that interference effects are only sizeable for
gluon-induced processes and lower the hadronic cross section by
about $5\,\%$. The dominant contributions result from interferences of
the QCD $\ttbarbbbar$ production diagrams of order
$\order{\alphas^2\alpha^2}$ with diagrams of order
$\order{\alphas\alpha^3}$ involving t-channel W-boson exchange.
Interferences between the dominant background and the $\ttbarh$ signal
on the other hand are below one per cent.  In most of the differential
distributions the interference effects lead to a constant shift. We
found, however, a few distributions where non-uniform shape
changes appear.

Our analysis demonstrates the complexity of this process and
provides useful information for a future NLO calculation. We found
that the \ttbarbbbar process provides a good approximation to the full
process, with a deviation of only about $8\,\%$ for the total cross
section and a uniform shift for most differential distributions. While
the QCD corrections to the leading QCD contributions to $\ttbarbbbar$
production are already known, a calculation of the NLO corrections to
the full $\ttbarbbbar$ process should be feasible with available tools
\cite{Actis:2012qn,Denner:2014gla}.

\acknowledgments 
This work was supported by the Bundesministerium
f\"ur Bildung und Forschung (BMBF) under contract no. 05H12WWE.

\appendix

\section{On-shell projection}
\label{sec:OnShellProjection}

In this appendix we discuss some details of our implementation of the pole 
approximation. For \ttbarbbbar production (\ttbarh production) we compute the 
matrix element with on-shell momenta for the top quarks (and the Higgs boson) 
and consider off-shell effects only in the corresponding propagators of the 
unstable particles. Thus, starting from off-shell momenta we use an on-shell 
projection to generate the momenta of the resonant particles. Since the 
procedure of on-shell projection is not uniquely defined we impose suitable 
requirements. The matrix element is strongly sensible to resonances and hence we 
project such as to retain sensible invariants as far as possible. As a 
consequence of the on-shell projection of the top quarks we also need to 
incorporate proper on-shell projections for their decay products.

\subsection{Higgs-boson on-shell projection}
\label{sec:HiggsOnShellProjection}

For \ttbarh production we project the Higgs boson in $\PH\to 
\Pb\bar{\Pb}$ on-shell, such that $\hat{p}_\text{H}^2=m_\text{H}^2$,  
$\hat{p}_\Pb^2=p_\Pb^2=m_\Pb^2$, $\hat{p}_{\bar{\Pb}}^2=p_{\bar{\Pb}}^2=m_\Pb^2$ 
and $\hat{p}_\text{H}=\hat{p}_\Pb+\hat{p}_{\bar{\Pb}}$, where $\hat{p}$ denotes 
the projected momentum corresponding to $p$. With these constraints
the projected momenta are obtained as
\begin{equation}
        \begin{aligned}
                \hat{p}_\Pb         &= \alpha\,\frac{p_\Pb-p_{\bar{\Pb}}}{2}+\beta\, \frac{p_\Pb+p_{\bar{\Pb}}}{2},\\ 
                \hat{p}_{\bar{\Pb}} &= \alpha\,\frac{p_{\bar{\Pb}}-p_\Pb}{2}+\beta\, \frac{p_\Pb+p_{\bar{\Pb}}}{2}, \\
                \hat{p}_\PH         &= \beta p_\PH,
        \end{aligned}
\end{equation}
where
\begin{equation}
        \alpha = \sqrt{\frac{\MH^2-4\Mb^2}{\ph^2-4\Mb^2}},\qquad
        \beta  = \frac{\MH}{\sqrt{\ph^2}}.
\end{equation}
We choose the positive solution for $\alpha$ to minimise momenta alteration. 

To preserve momentum conservation we choose to adjust the top-quark momentum according to
\begin{equation}\label{eqn:TopQuarkAdjustment}
        \pt^\prime = \pt + \ph - \phath.
\end{equation}
Adjusting instead the antitop-quark momentum accordingly would lead to
a slightly different but equally well performing on-shell projection.

\subsection{Top- and antitop-quark on-shell projection}
\label{sec:tt-projection}

We project the top and the antitop quarks on-shell simultaneously. In
the case of the \ttbarh approximation the Higgs boson has been
projected on-shell (see above) and the top-quark momentum gets
adjusted \eqref{eqn:TopQuarkAdjustment} to preserve momentum
conservation. In the following we use the adjusted top-quark momentum
$\pt^\prime$ for \ttbarh production while for \ttbarbbbar production
the original momentum is used. The antitop-quark momentum is the
original one $\ptx$ for both the \ttbarh and the \ttbarbbbar scenario.

The imposed requirements for the simultaneous top and
antitop projection are the on-shell conditions $\phatt^2=\Mt^2$ and
$\phattx^2=\Mt^2$ and momentum conservation,
$\phatt+\phattx=\pt^\prime+\ptx$. We consider the general case of two
momenta $p_1$ and $p_2$ to be suitably projected.
We define $q=p_1+p_2$ and seek to construct projected momenta $\hat
p_1$ and $\hat p_2$ fulfilling $\hat{p}_1+\hat{p}_2=q$ and
$\hat{p}_1^2$ and $\hat{p}_2^2$ fixed by given values not necessarily
the on-shell conditions $\hat{p}_1^2=m_1^2$ and $\hat{p}_2^2=m_2^2$.
With the ansatz
\newcommand{\ptprime}{\pt^\prime}
\newcommand{\ptprimesq}{\pt^{\prime\,2}}
\begin{equation}
                \hat{p}_1 =    \xi  p_1 +    \eta  p_2, \qquad
                \hat{p}_2 = (1-\xi) p_1 + (1-\eta) p_2
\end{equation}
we obtain
\begin{equation}\label{eqn:OnShellProjectionXi}
                \xi = \frac{\left(\left(p_1+p_2\right)^2+\hat{p}_1^2-\hat{p}_2^2\right)-2\eta \left(p_2^2+p_1 p_2\right)}{2\left(p_1^2+p_1 p_2\right)}
\end{equation}
and the quadratic equation for the determination of $\eta$:
\begin{equation}
        \begin{aligned}\label{eqn:OnShellProjectionEta}
                0 =\, & \eta^2\left[p_1^2p_2-p_2^2p_1+(p_1p_2)(p_2-p_1)\right]^2\\
                        &+ \eta\left[\left(p_1+p_2\right)^2+\hat{p}_1^2-\hat{p}_2^2\right]\left[(p_1 p_2)^2-p_1^2 p_2^2\right] \\
                        &+ \frac{1}{4}\left[\left(p_1+p_2\right)^2+\hat{p}_1^2-\hat{p}_2^2\right]^2 p_1^2-\left(p_1^2+p_1 p_2\right)^2\hat{p}_1^2.
        \end{aligned}
\end{equation}
We choose the smaller solution for $\eta$ to minimise momenta alteration.
For the on-shell projection of the top quark and antitop quark we identify 
$p_1\to\ptx$, $p_2\to \pt^\prime$, analogously for the projected momenta, and 
$\hat{p}_1^2=\hat{p}_2^2=\Mt^2$ .

\subsection{On-shell projection of the top- and antitop-quark decay products}

The on-shell projection of the top and antitop quark alters the
momenta of their decay products, which have to be adjusted
accordingly. We compute new momenta of the bottom ($\pbtprime$) from
the top-quark decay ($\Pt \to \PW^+ \Pb$) and antibottom
($\pbxtxprime$) from the antitop-quark decay ($\bar{\Pt} \to \PW^-
\bar{\Pb}$), respectively, via
\begin{equation}
                \pbtprime   = \phatt  - \pwplus, \qquad
                \pbxtxprime = \phattx - \pwminus
\end{equation}
and project them on-shell. To this end we project the $\PW^+$ and b 
simultaneously with the requirements $\phatwplus+\phatbt = \pwplus+\pbtprime$, 
$\phatbt^2=\Mb^2$ and $\phatwplus^2=\pwplus^2$. The latter condition preserves 
the resonance of the $\PW^+$ propagator, since the $\PW^+$ is likely to be 
nearly on-shell and an alteration of its propagator is undesirable. We use a 
similar projection as in \refse{sec:tt-projection}, i.e.\
\begin{equation}
        \begin{aligned}
                \phatwplus &=    \tilde\xi  \pwplus +    \tilde\eta  \pbtprime, \\
                \phatbt    &= (1-\tilde\xi) \pwplus + (1-\tilde\eta) \pbtprime
        \end{aligned}
\end{equation}
with \eqref{eqn:OnShellProjectionXi} and \eqref{eqn:OnShellProjectionEta} and 
the identification $p_1\to\pwplus$, $\hat{p}_1^2\to\pwplus^2$, 
$p_2\to\pbtprime$, and $\hat{p}_2^2\to \Mb^2$. Analogously we project $\PW^-$ 
and $\bar{\Pb}$ of the decay $\bar{\Pt}\to \PW^-\bar{\Pb}$.

Two more projections are required for the decays $\PW^+\to \Pl^+\nu_\Pl$ and 
$\PW^-\to \bar{\Pu}\Pd$ (or $\PW^-\to \bar{\Pc}\Ps$) to ensure the on-shellness 
of the final-state particles. We explain the procedure by the projection of the 
charged lepton and neutrino.  We first redefine the neutrino momentum via
\begin{equation}
 p'_{\nu_\Pl} = \phatwplus  - p_{\Pl^+}
\end{equation}
to restore momentum conservation. Then, we make the ansatz
\begin{equation}\label{eqn:MomentumModification}
                \hat{p}_{\Pl^+} = \alpha p_{\Pl^+}, \qquad
                \hat{p}_{\nu_\Pl} = \phatwplus  - \hat{p}_{\Pl^+}
\end{equation}
that trivially fulfils the on-shell condition for the charged lepton, $\hat 
p_{\Pl^+}^2=0$. Requiring the projected neutrino momentum to be on-shell yields 
the following on-shell projected momenta of the $\PW^+$ decay products in 
$\PW^+\to \Pl^+\nu_\Pl$:
\begin{equation}
                  \hat{p}_{\nu_\Pl} =  p'_{\nu_\Pl} -
                  \frac{p_{\nu_\Pl}^{\prime 2}}{2p'_{\nu_\Pl} p_{\Pl^+}} p_{\Pl^+},\qquad
                  \hat{p}_{\Pl^+} = \left(1+\frac{p_{\nu_\Pl}^{\prime 2}}{2p'_{\nu_\Pl} p_{\Pl^+}}\right)  p_{\Pl^+}.
\end{equation}
Analogously we obtain for the projection of the $\PW^-$ decay products in $\PW^-\to 
\bar{\Pu}\Pd(\bar{\Pc}\Ps)$:
\begin{equation}
                  \hat{p}_{\Pd(\Ps)} =  p'_{\Pd(\Ps)} -
                  \frac{p_{\Pd(\Ps)}^{\prime 2}}{2p'_{\Pd(\Ps)} p_{\bar{\Pu}(\bar{\Pc})}}p_{\bar{\Pu}(\bar{\Pc})},\qquad
                  \hat{p}_{\bar{\Pu}(\bar{\Pc})} =
                  \left(1+\frac{p_{\Pd(\Ps)}^{\prime 2}}{2p'_{\Pd(\Ps)} p_{\bar{\Pu}(\bar{\Pc})}}\right) p_{\bar{\Pu}(\bar{\Pc})}
\end{equation}
with 
\begin{equation}
 \hat{p}'_{\Pd(\Ps)} = \phatwminus  - \hat{p}_{\bar{\Pu}(\bar{\Pc})}.
\end{equation}
Thus, we perform six projections in total for \ttbarh production and five for 
\ttbarbbbar production. Of course, the on-shell projections do not
change the momenta, if the resonant particles are already on shell.

\section{Further differential distributions}
\label{sec:FurtherObs}
In this appendix we present further differential distributions which
exhibit only minor shape deviations between the full process and the
approximations. As for the total cross
section we find for most distributions a constant offset of $8\,\%$ between the full process and the
$\ttbbb$ approximation and about a factor four between the full
process and the $\ttbarh$ approximation.  Only in tails of
distributions larger deviations between the full
process and the approximations appear.

\begin{figure}
        \setlength{\parskip}{-10pt}
        \begin{subfigure}{0.48\linewidth}
                \subcaption{}
                \label{plot:transverse_momentum_positron}
                \includegraphics[width=\linewidth]{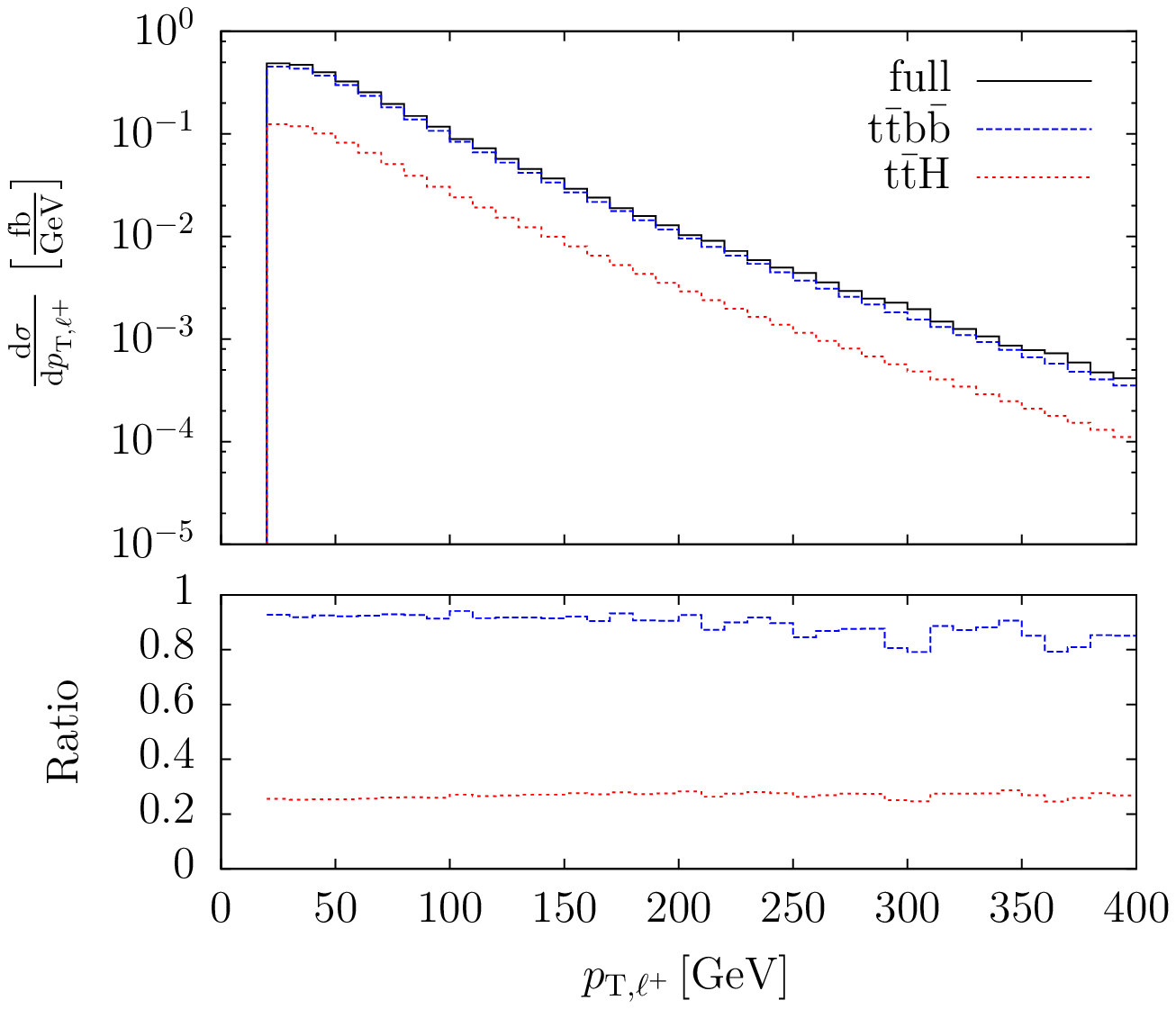}
        \end{subfigure}
        \hfill
        \begin{subfigure}{0.48\linewidth}
                \subcaption{}
                \label{plot:rapidity_positron}
                \includegraphics[width=\linewidth]{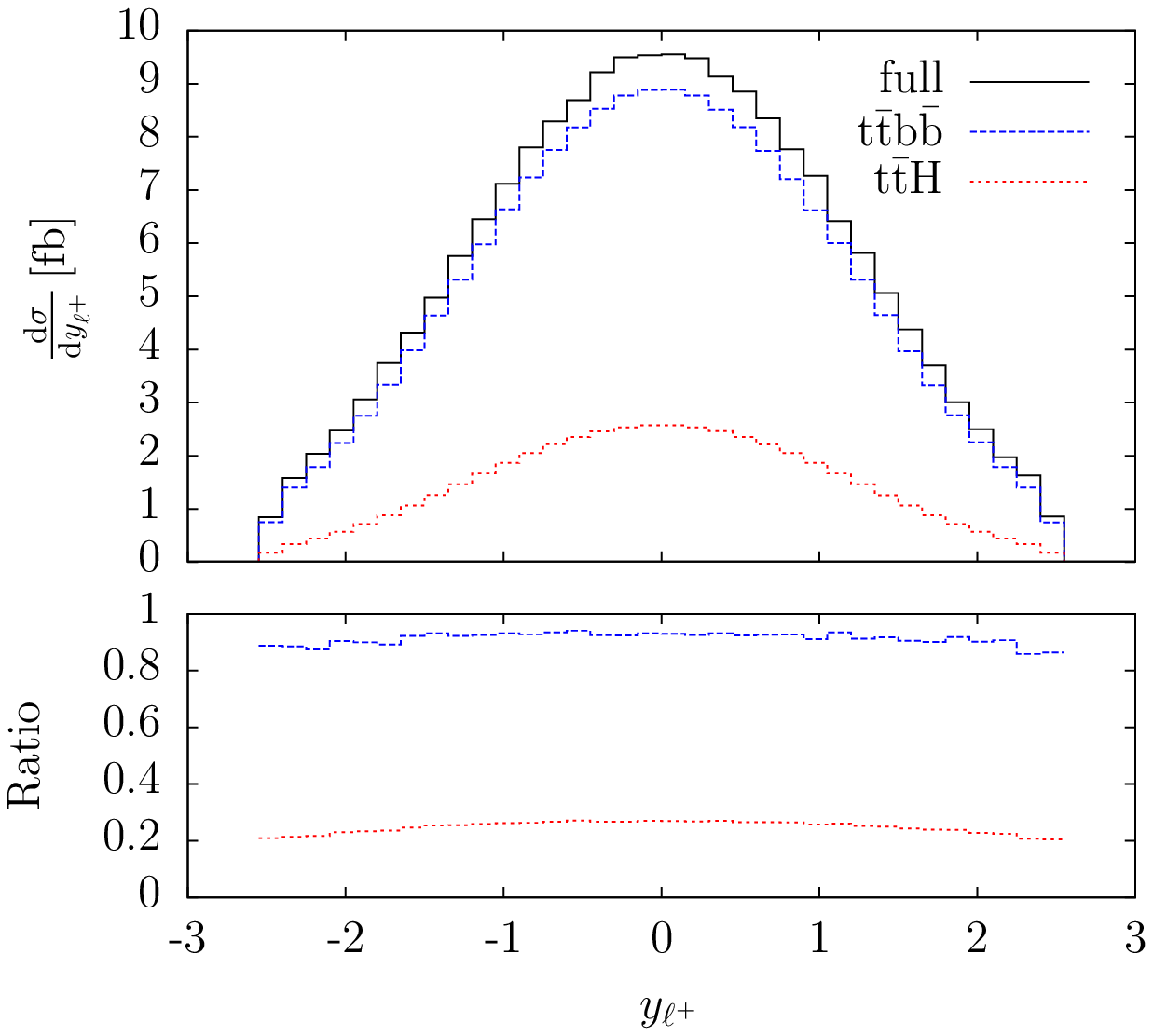}
        \end{subfigure}
        \begin{subfigure}{0.48\linewidth}
                \subcaption{}
                \label{plot:transverse_momentum_hardest_bjet}
                \includegraphics[width=\linewidth]{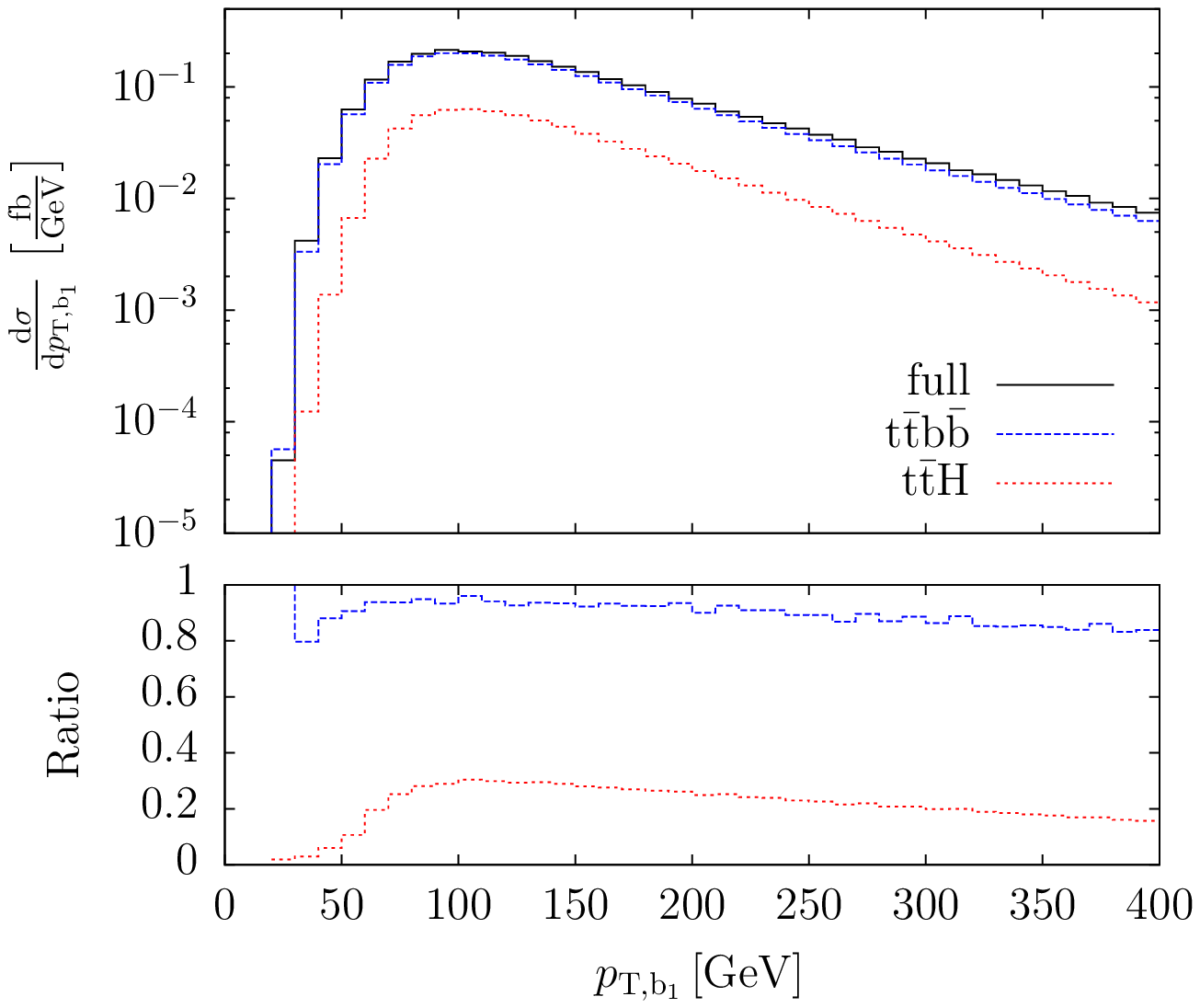}
        \end{subfigure}
        \hfill
        \begin{subfigure}{0.48\linewidth}
                \subcaption{}
                \label{plot:rapidity_hardest_bjet}
                \includegraphics[width=\linewidth]{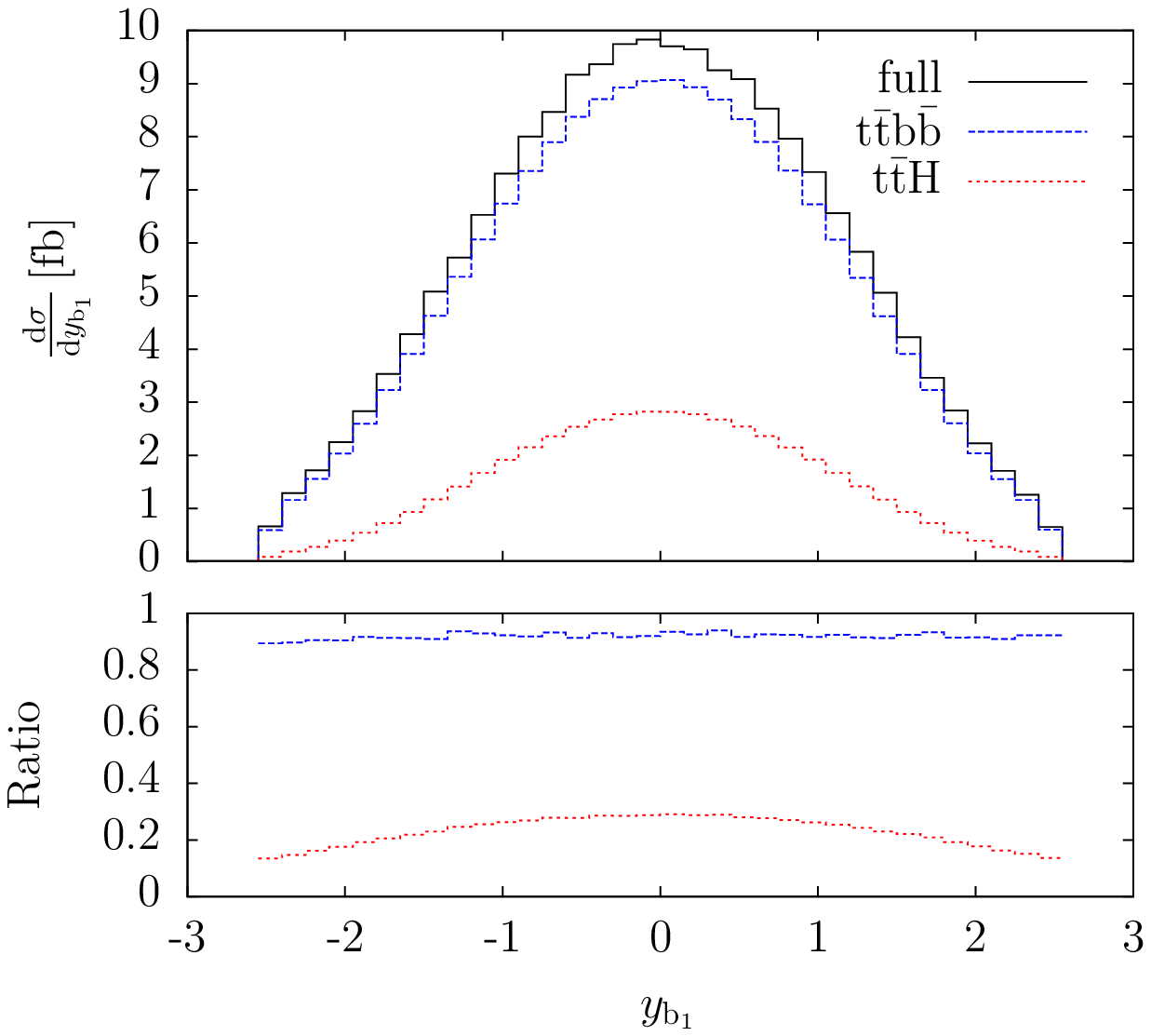}
        \end{subfigure}
        \begin{subfigure}{0.48\linewidth}
                \subcaption{}
                \label{plot:missing_transverse_momentum} 
                \includegraphics[width=\linewidth]{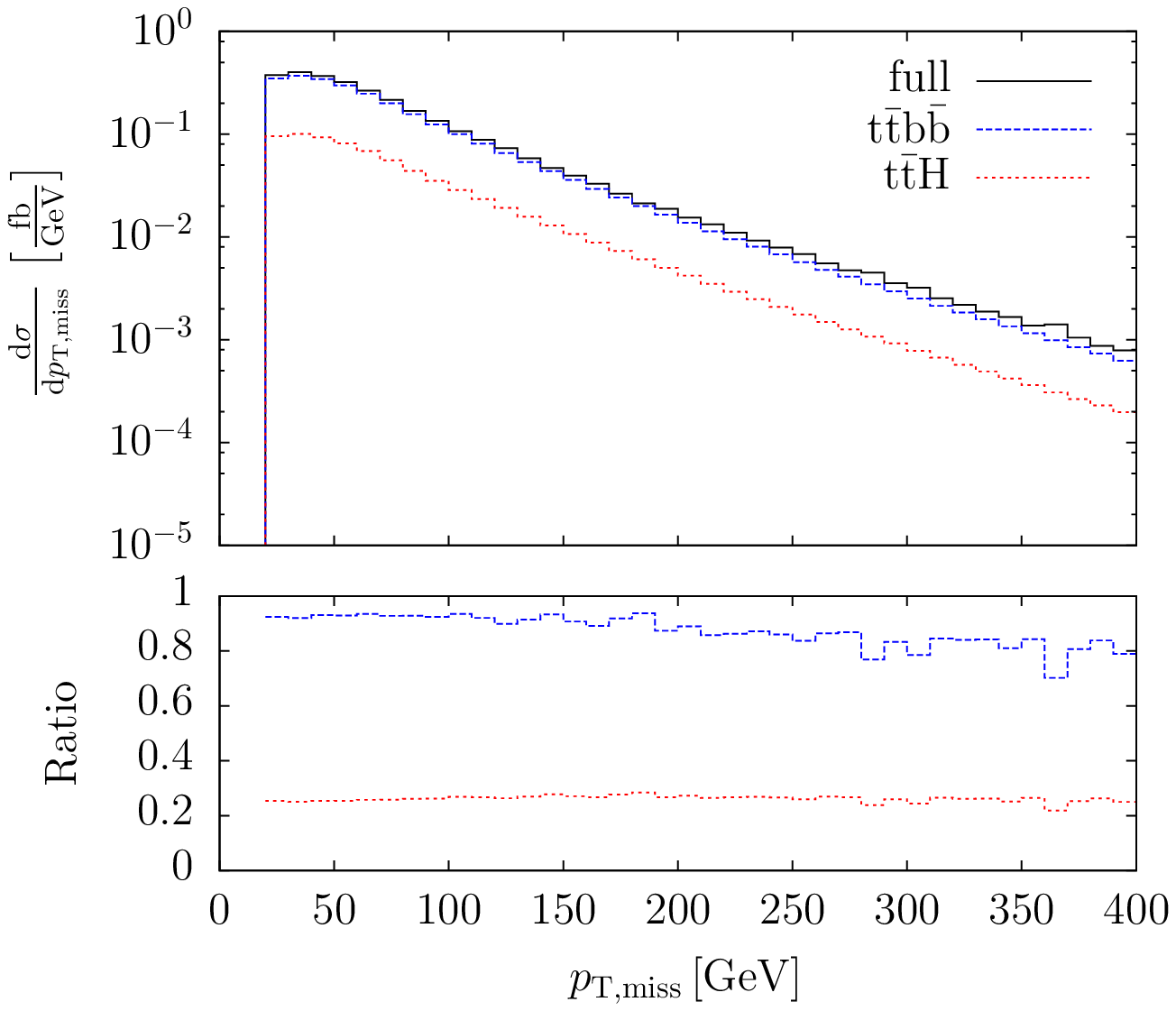}
        \end{subfigure}
        \hfill
        \begin{subfigure}{0.48\linewidth}
                \subcaption{}
                \label{plot:transverse_mass_epveb1_born_lo}
                \includegraphics[width=\linewidth]{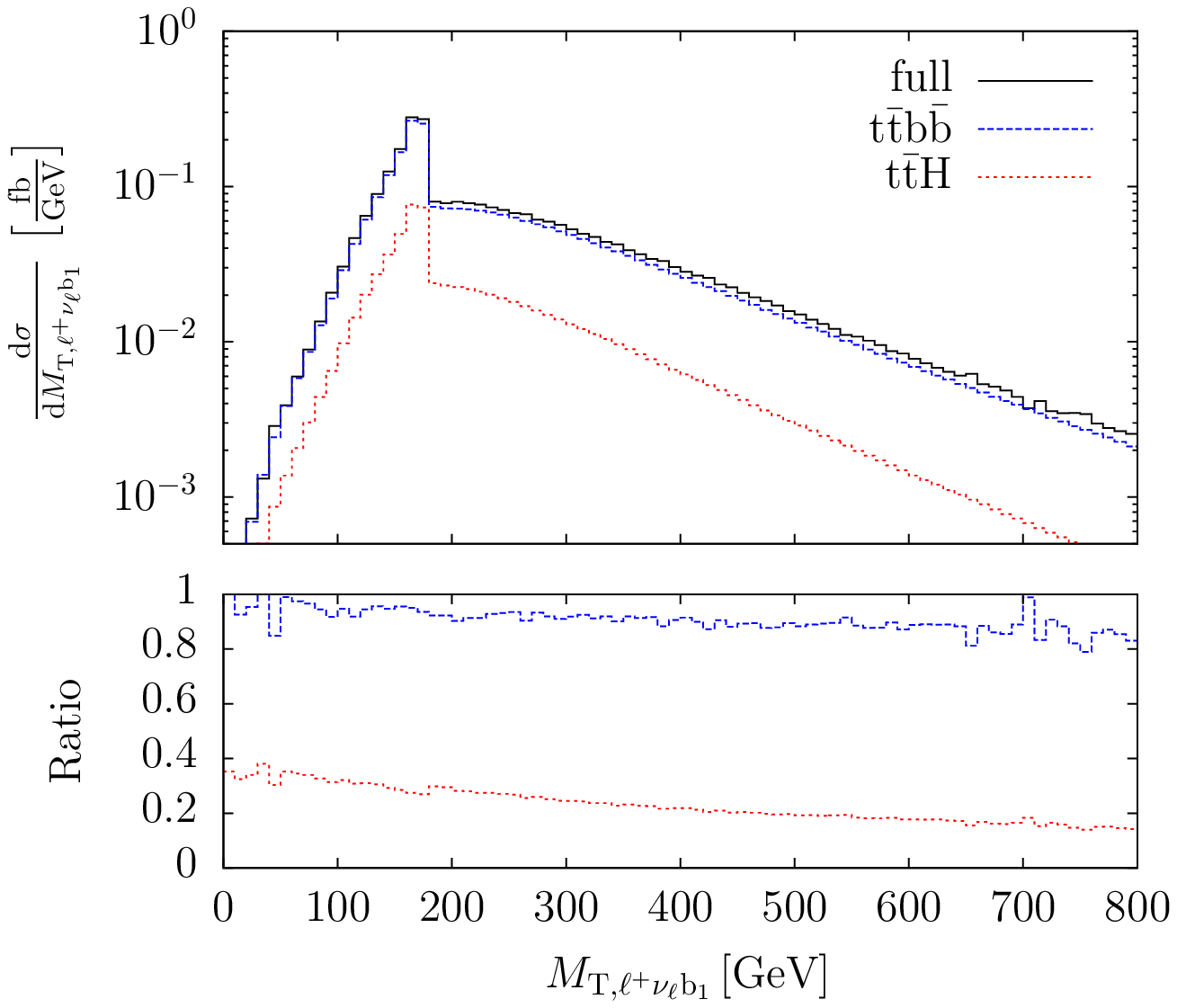}
        \end{subfigure}
        \caption{Standard differential distributions at the LHC at $13\TeV$ for the full process, 
        \ttbarbbbar production and \ttbarh production:     
        \subref{plot:transverse_momentum_positron}~transverse momentum of the charged lepton (upper left), %
        \subref{plot:rapidity_positron}~rapidity of the charged lepton (upper right), %
        \subref{plot:transverse_momentum_hardest_bjet}~transverse momentum of the hardest \Pb~jet (middle left), %
        \subref{plot:rapidity_hardest_bjet}~rapidity of the hardest \Pb~jet (middle right), %
        \subref{plot:missing_transverse_momentum}~missing transverse momentum (lower left), %
        \subref{plot:transverse_mass_epveb1_born_lo}~transverse mass of the $\Pl^+\nu_\Pl \Pb_1$ system (lower right right). %
        The lower panels show the relative size of $\ttbbb$ and $\ttbarh$ production normalised to the full process.}
\end{figure}

In
\reffis{plot:transverse_momentum_positron}--\ref{plot:rapidity_positron}
we show the transverse momentum and the rapidity distribution of the
identified charged lepton, while the transverse momentum and the
rapidity distribution of the hardest b~jet are depicted in
\reffis{plot:transverse_momentum_hardest_bjet}--\ref{plot:rapidity_hardest_bjet}.
Finally, the missing transverse momentum and the transverse mass of
the charged lepton, the neutrino and the hardest b~jet are provided in
\reffis{plot:missing_transverse_momentum}--\ref{plot:transverse_mass_epveb1_born_lo}.

\bibliographystyle{jhep}
\bibliography{ttxh_lo} 

\end{document}